\documentclass[reprint]{revtex4-1}

\usepackage{graphicx}
\usepackage{dcolumn}
\usepackage{bm}

\usepackage[utf8]{inputenc}
\usepackage[T1]{fontenc}
\usepackage{mathptmx}
\usepackage{etoolbox}
\usepackage{amsmath}

\makeatletter
\def\@email#1#2{%
 \endgroup
 \patchcmd{\titleblock@produce}
  {\frontmatter@RRAPformat}
  {\frontmatter@RRAPformat{\produce@RRAP{*#1\href{mailto:#2}{#2}}}\frontmatter@RRAPformat}
  {}{}
}%
\makeatother
\begin{document}

\preprint{AIP/123-QED}

\title{Rotationally inelastic scattering of cyanocyclopentadiene by helium atoms}

\author{Karina Sogomonyan}
\affiliation{KU Leuven, Department of Chemistry, Celestijnenlaan 200F, 3001 Leuven, Belgium.}

 \email{karina.sogomonyan@kuleuven.be}
 
\author{Malek Ben Khalifa}%
 \affiliation{KU Leuven, Department of Chemistry, Celestijnenlaan 200F, 3001 Leuven, Belgium.}

\author{Phoebe Pierré}
 \affiliation{KU Leuven, Department of Chemistry, Celestijnenlaan 200F, 3001 Leuven, Belgium.}

\author{Jérôme Loreau}
 \email{jerome.loreau@kuleuven.be}
\affiliation{KU Leuven, Department of Chemistry, Celestijnenlaan 200F, 3001 Leuven, Belgium.}%

\date{\today}

\begin{abstract}
In the interstellar medium (ISM), polycylic aromatic hydrocarbons (PAHs) are believed to be an important carbon reservoir, accounting for up to a quarter of all interstellar carbon in our galaxy. This makes the investigation of their potential formation precursors highly relevant in the context of ISM chemistry. This, in turn, includes knowing the abundance of the precursor species. One of the possible precursor molecules for PAHs is the recently detected cyanocyclopentadiene, c-C$_5$H$_5$CN. Given the physical conditions of the dense dark molecular cloud TMC-1 where the cyclic species was detected, it is crucial to consider that local thermodynamic equilibrium conditions may not be satisfied. In such case an accurate estimation of the molecular abundance involves taking into account the competition between the radiative and collisional processes, which requires the knowledge of rotational excitation data for collisions with the most abundant interstellar species - He or H$_2$. In this paper the first potential energy surface (PES) for the interaction of the most stable isomer of cyanocyclopentadiene (1-cyano-1,3-cyclopentadiene) with He atoms is computed using the explicitly correlated coupled-cluster theory [CCSD(T)-F12]. The obtained PES demonstrates a high anisotropy and is characterized by a global potential well of -101.8 cm$^{-1}$. Scattering calculations of the rotational (de-)excitation of 1-cyano-cyclopentadiene induced by He atoms are performed with the quantum mechanical close-coupling method for total energies up to 125 cm$^{-1}$. The resulting rotational state-to-state cross sections are used to compute the corresponding rate coefficients for temperatures up to 20 K. A propensity favoring the transitions with $\Delta k_a=0$ is observed.
\end{abstract}

\maketitle

\section{Introduction} {\label{introduction}}
Polycyclic aromatic hydrocarbons (PAHs) are carbon-based compounds consisting of two or more fused aromatic rings. PAHs are ubiquitous on earth: they are usually found in the atmosphere due to being the by-products of fossil fuel combustion \cite{yamamoto2017introduction}.
In the interstellar medium (ISM) PAHs are recognized to be the carriers of the unidentified IR (UIR) bands \cite{tielens2008interstellar}. While the UIR bands are not assigned to specific molecules, they indicate the presence of PAHs in a wide number of sources in the ISM. Thus, PAHs are estimated to allocate up to $10-25\%$ of all interstellar carbon in the Milky Way \cite{wenzel2024detection}.
Their detection is usually hindered due to the absence of strong dipole moments or lack of the laboratory data. However, several PAHs such as 1- and 2-cyanonaphthalene \cite{mcguire2021detection}, pure indene \cite{burkhardt2021discovery}, 2-cyanoindene \cite{sita2022discovery}, 1-cyanopyrene \cite{wenzel2024detection}, and two derivatives of acenaphtylene (1- and 5-cyanonaphtylene) \cite{cernicharo2024discovery} have been recently observed in the ISM.\\
Formation pathways of PAHs are still a subject of debate. Two contrasting mechanisms are suggested: the "top-down" scenario proposes PAH formation through the destruction of larger dust grains by harsh radiation \cite{tielens2008interstellar,refId0}, whereas the "bottom-up" approach suggests that complex molecules are synthesized from smaller carbon-containing precursors in the cold environments of dark clouds \cite{annurev:/content/journals/10.1146/annurev-physchem-040214-121502}. The second approach explains the presence of small PAHs (less than 35 atoms) in the ISM where they were believed to be destroyed according to previous models \cite{wenzel2024detection}. This highlights the importance of investigating the precursors of PAH formation: single six- and five-membered rings. Knowing the abundance of cyclic precursors of PAHs in astronomical environments can provide insight into possible reaction pathways. However, pure cycles present an observational challenge for radio astronomy due to the lack of a permanent dipole moment. Consequently, the first detection of an unsubstituted aromatic molecule --  benzene (C$_6$H$_6$) was achieved through its weak infrared absorption feature attributed to the $v_4$ bending mode in the proto-planetary nebula CRL 618 \cite{Cernicharo_2001}. Alternatively, functionalized cyclic molecules with large dipole moments produce more distinct rotational lines than their unsubstituted counterparts. Cyano-containing aromatics are formed through fast barrierless reactions under typical interstellar conditions and can be used as an observational proxy for the presence of the aromatic rings \cite{Cooke_2020}. Considering this, CN-substituted species have become a subject of interest resulting in multiple detections in the recent years: benzonitrile (C$_6$H$_5$CN) \cite{mcguire2018detection}, and two isomers of cyanocyclopentadiene --  1-cyano-1,3-cyclopentadiene (1-cyano-CPD, 1-C$_5$H$_5$CN) \cite{mccarthy2021interstellar} and 2-cyano-1,3-cyclopentadiene (2-cyano-CPD, 2-C$_5$H$_5$CN) \cite{lee2021interstellar} were observed toward TMC-1. The latter detections were followed by the first detection of a pure five-membered ring molecule cyclopentadiene in the same source \cite{cernicharo2021pure}. This opened the possibility of directly linking cyclic molecules with their functionalized nitrile derivatives in a cyano/pure ring relationship, further supporting the validity of using cyano- derivatives as proxies for their pure cyclic analogues \cite{sita2022discovery}. \\
To estimate molecular abundances in low-temperature environments (e.g., TMC-1) it is common to assume Local Thermodynamic Equilibrium (LTE) conditions and a single excitation temperature \cite{mccarthy2021interstellar,cernicharo2021discovery}. This assumption is generally accepted for TMC-1, yet it has not been explicitly verified for complex organic molecules (COMs). Moreover, the recent results of a simple radiative transfer calculation show that benzonitrile is likely not at LTE in the cold clouds \cite{ben2024rotational}. In this case, precise collisional excitation rate coefficients are necessary to account for the competition between the radiative and collisional processes. Consequently, a collisional excitation study of cyano-CPD, another simple organic cycle detected in the same region, could provide a better understanding of cyclic organic molecules in cold molecular clouds.\\
Cyano-CPD exists in the form of three isomers: 1-cyano-CPD, 2-cyano-CPD, and 5-cyano-CPD. Their stability was quantified at the G3//B3LYP level of theory, determining 1-cyano-CPD to be the most stable isomer, followed by 2-cyano-CPD (5 kJ/mol higher) and 5-cyano-CPD (26 kJ/mol higher) \cite{mccarthy2021interstellar}. Possible formation pathways for the two most stable cyano-CPD isomers have also been investigated at the same level of theory. The results show that both 1-cyano-CPD and 2-cyano-CPD are formed through a  typical neutral-neutral reaction of cyclopentadiene with CN radical. The reported reaction pathway is exothermic and barrierless irrespective of the relative stability of the two isomers. The column densities derived from observations show that 1-cyano-CPD is more abundant than 2-cyano-CPD by a factor of 3-4 \cite{lee2021interstellar}. It was suggested that not only the difference in stability might be the cause for such discrepancy, but an additional formation mechanism for 1-cyano-CPD through reaction with butadiene: CCN +  CH$_2$CHCHCH$_2$ $\xrightarrow{}$ c-C$_5$H$_5$CN (1-cyano-CPD) + H \cite{cernicharo2021discovery}.\\
Collisional excitation studies of large COMs present various computational challenges. Increasing molecular complexity results in strong anisotropy of the corresponding interaction potential energy surface (PES). Additionally, a high density of rotational states results in a large number of rotational levels populated at low energy, with consequent full-quantum scattering calculations approaching the current computational limit. Generally, the collisional properties of interest are those involving the most abundant interstellar species, such as H$_2$, as a collider. Still, due to the exceptional computational demands for COM-H$_2$ collisions, He is often used as a proxy for para-H$_2$ ($j=0$), while taking the mass difference into account by scaling the rate coefficients. To this day, only several cyclic COM species have been studied in collision with He: propylene oxide \cite{dzenis2022collisional} and cyclopentadiene \cite{demes2024first} with the close coupling (CC) method, benzonitrile using the coupled states (CS) approximation \cite{ben2024rotational}, and benzene by means of Mixed Quantum/Classical Theory (MQCT) \cite{mandal2022mixed}. \\
In this work we present scattering cross sections and rate coefficients for collisional (de-)excitation of 1-cyano-CPD with He atoms at low temperature based on a first accurate potential energy surface and fully quantum scattering calculations. In section \ref{PES_section} we present the \textit{ab initio} study of the 1-cyano-CPD -- He interaction and the construction of the associated PES. In section \ref{dynamics} we report the scattering dynamics and illustrate the inelastic cross sections for 1-cyano-CPD collisions with He. Collisional quenching rate coefficients derived for kinetic temperatures up to 20 K are presented in section \ref{ratecoefficients}. We discuss conclusions and outlook in section \ref{conclusions}.\\

\section{Potential Energy Surface} \label{PES_section}
\subsection{\textit{Ab initio} calculations} 
\begin{figure}[h]
    \centering
    \includegraphics[width=0.95\linewidth]{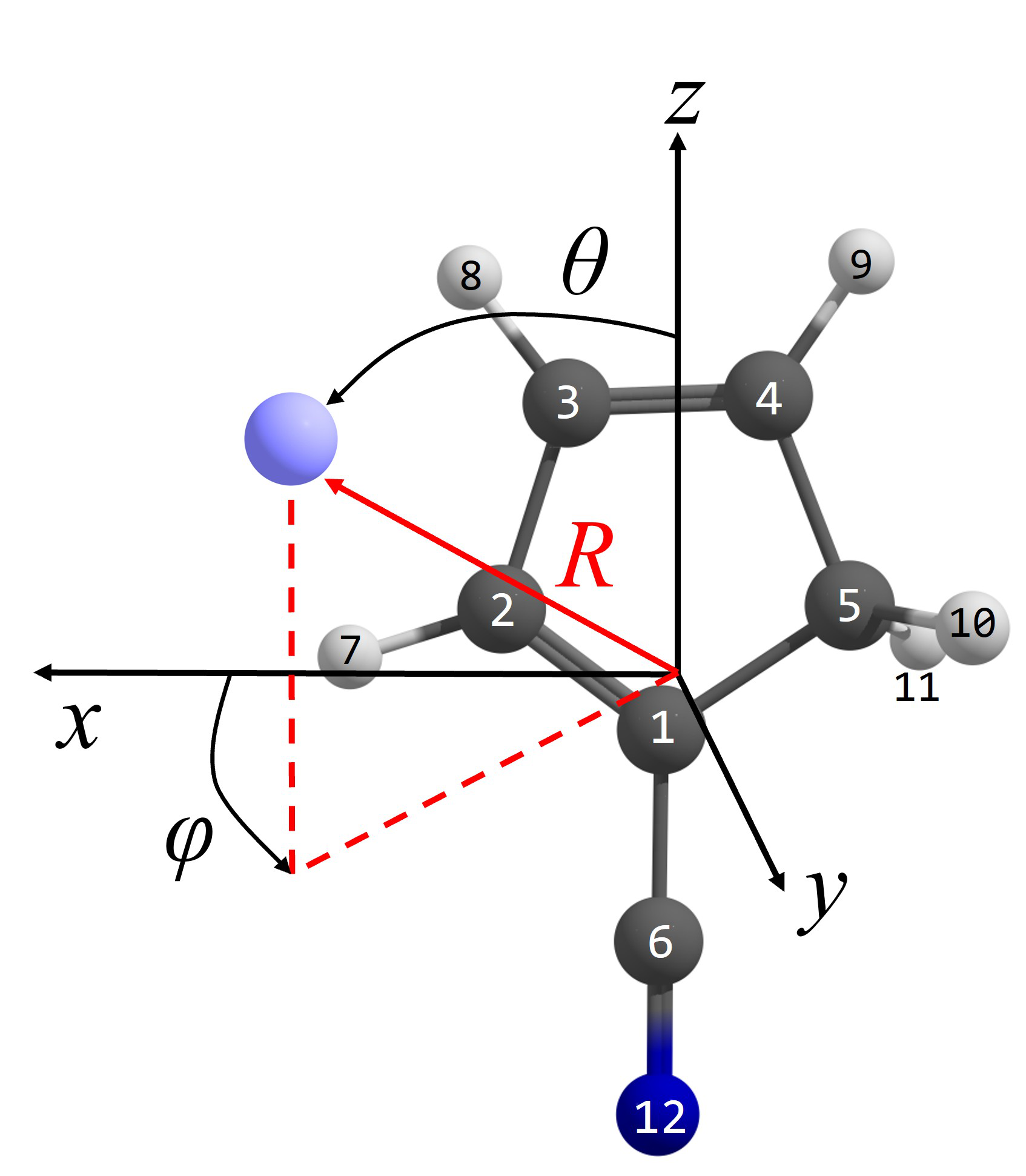}
    \caption{Jacobi coordinate system for the 1-cyano-CPD-He complex. The coordinate system origin is in the center of mass of 1-cyano-CPD.}
    \label{fig:coordsystem}
\end{figure}

\begin{figure}
    \centering
    \includegraphics[width=1\linewidth]{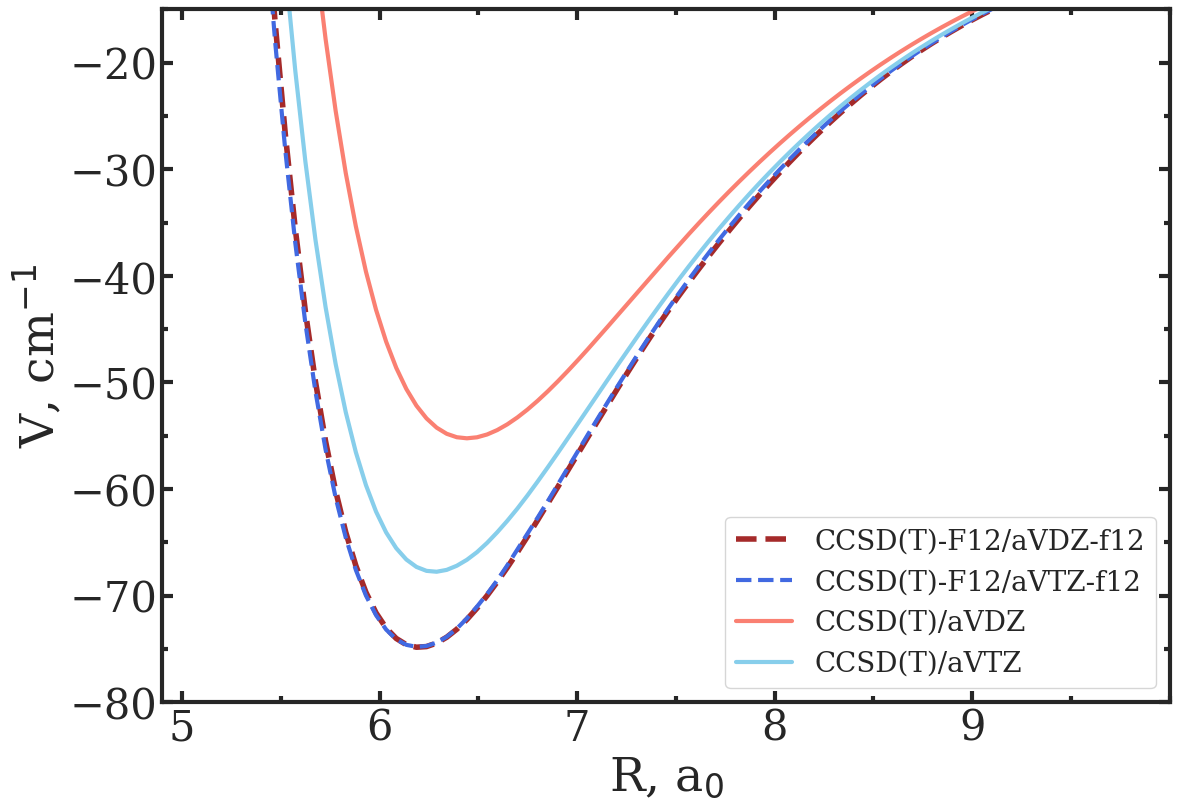}
    \caption{Potential energy surface cuts (cm$^{-1}$) for 1-cyano-CPD -- He complex as a function of $R$ for $\phi=90^{\circ}$, $\theta=90^{\circ}$.}
    \label{fig:basis-sets}
\end{figure}

We start by computing the PES of the interaction between the asymmetric top molecule 1-cyano-CPD and a helium atom in their ground states using the rigid-rotor approximation. Preliminary calculations with previously derived geometry parameters \cite{sakaizumi1987microwave} for 1-cyano-CPD resulted in significant deviations of rotational constants from those reported \cite{ford1978microwave}. Hence, the geometry parameters used in PES construction were first optimized at the CCSD(T)-F12/aug-cc-pVTZ level of theory providing the best match for the rotational constants. The optimized internal coordinates are reported in Table \ref{geom} using the numbering of the atoms shown in Figure \ref{fig:coordsystem}.\\
\begin{table} 

\caption{Equilibrium geometry of 1-cyano-CPD optimized at the CCSD(T)-F12/aug-cc-pVTZ level of theory}
\label{geom}
\begin{ruledtabular}
\begin{tabular}{cccc}
 &$r$ (\AA)& & angle $(^{\circ})$\\
\hline
$r$(C$_1$--C$_2$)& 1.352 & $\angle$(C$_2$C$_1$C$_5$) & 109.5  \\
$r$(C$_2$--C$_3$)& 1.455 & $\angle$(C$_1$C$_2$C$_3$) & 109.3 \\
$r$(C$_3$--C$_4$) & 1.349 & $\angle$(C$_2$C$_3$C$_4$) & 109.2 \\
 $r$(C$_4$--C$_5$) & 1.504 & $\angle$(C$_3$C$_4$C$_5$) & 109.7 \\
 $r$(C$_1$--C$_5$) & 1.504 & $\angle$(C$_4$C$_5$C$_1$) & 102.4 \\
 $r$(C$_1$--C$_6$) & 1.420 & $\angle$(H$_{10}$C$_5$H$_{11}$) & 107.4 \\
 $r$(C$_6$$\equiv$N$_{12}$) & 1.162 & $\angle$(C$_1$C$_5$H$_{11}$) & 111.7 \\
 $r$(C$_2$--H$_7$) & 1.086 & $\angle$(C$_3$C$_2$H$_7$) & 125.0 \\
 $r$(C$_3$--H$_8$) & 1.080 & $\angle$(C$_2$C$_3$H$_8$) & 124.5 \\
 $r$(C$_4$--H$_9$) & 1.080 & $\angle$(C$_5$C$_4$H$_9$) & 123.9 \\
 $r$(C$_5$--H$_{10}$) & 1.094 & & \\
 $r$(C$_5$--H$_{11}$) & 1.094 & & \\
\end{tabular}
\end{ruledtabular}
\end{table}
The Jacobi coordinates ($R$, $\theta$, $\phi$) defining the interaction between the collision partners are shown in Figure \ref{fig:coordsystem}. $R$ represents the magnitude of the vector $\mathbf{R}$ connecting the mass center of 1-cyano-CPD and the helium atom, $\theta$ and $\phi$ are the polar and azimuthal angles. The axes $x$, $y$ and $z$ are aligned with the inertia axes $b$, $c$ and $a$ respectively, the origin of the axes is the mass center of 1-cyano-CPD and the molecule lies in the $xz$ plane.\\
All \textit{ab initio} calculations for geometry optimization and PES construction were carried out with the MOLPRO (version 2020.2) \cite{10.1063/5.0005081} program package. The interaction potential is defined using the counterpoise correction scheme of Boys and Bernardi \cite{boys1970calculation}.
Before the construction of the complete potential energy surface, we carried out an assessment of the method/basis set combination performance. Several potential energy surface cuts for different orientations of the helium atom with respect to the 1-cyano-CPD molecule were calculated with a standard CCSD(T) \cite{DEEGAN1994321} and explicitly correlated CCSD(T)-F12a method with single, double and perturbative triple excitations \cite{adler2007simple}. The augmented correlation-consistent basis sets (denoted as aug-cc-pVnZ, or aVnZ, where $n$= D, T) developed by Dunning et al. \cite{10.1063/1.456153} were used in these calculations. The CCSD(T)-F12/aVTZ method is considered to be the "gold standard" for the construction of a high precision global PES for the study of inelastic collisions. Due to the computational limitations we could not extend our preliminary study to the aVQZ basis set, thus considering the curves obtained at the CCSD(T)-F12/aVTZ level of theory to be the most accurate representation of the PES. Figure \ref{fig:basis-sets} displays a potential energy cut at a fixed geometry $\phi=90^{\circ}$, $\theta=90^{\circ}$ representing the approach of He atom perpendicular to the molecular plane of 1-cyano-CPD. As can be seen from the plot, CCSD(T)-F12/aVTZ and CCSD(T)-F12/aVDZ methods lead to almost identical results. 
After performing this assessment across multiple molecular orientations, the CCSD(T)-F12/aVDZ method was chosen for the global PES construction. This will yield an accurate PES while significantly reducing the disk space and CPU times. 
 \\
A total of 31958 \textit{ab initio} points were calculated to obtain the full potential energy surface. The chosen geometries represent 58 intermolecular distances from 4 a$_0$ to 50 a$_0$, 29 angles in $\theta$ varying between 0$^{\circ}$ and 180$^{\circ}$, and 19 values in $\phi$ in the range of 0$^{\circ}$ and 180$^{\circ}$ by steps of 10$^{\circ}$. A large number of points was necessary for a correct description of a highly anisotropic PES. To account for the size inconsistency of the explicitly correlated CCSD(T)-F12 methods \cite{knizia2009simplified} the entire PES was adjusted by the average asymptotic value of the interaction potential determined as -20.069 cm$^{-1}$ at $R$ = 100 a$_0$.
\subsection{Analytical Fit} \label{fit-subsect}
An expansion of the angular dependence of the PES over spherical harmonics is required to implement the interaction potential in quantum scattering calculations. Considering the symmetry of spherical harmonics, this expansion is written as: 
 \begin{equation}\label{eq_fit}
V(R,\theta,\phi)=\sum_{l=0}^{l_{\max}}\sum_{m=0}^{l}v_{lm}(R)\frac{Y_{l}^{m}(\theta,\phi)+(-1)^{m}Y_{l}^{-m}(\theta,\phi)}{1+\delta_{m,0}}
\end{equation}
where $v_{lm}(R)$ and $Y_{l}^{m}(\theta,\phi)$ correspond to the radial coefficients to be computed and the normalized spherical harmonics respectively, and $\delta_{m,0}$ is the Kronecker symbol. Continuous expansion coefficients $v_{lm}(R)$ were obtained through a least-squares for for every intermolecular distance for each value of the integers $l$ and $m$.\\
From the PES evaluated at 29 $\theta$ and 19 $\phi$ angles we were able to obtain radial coefficients up to $l_{\max}=18$, $m_{\max}=18$, resulting in 190 expansion terms. To assess the quality of the fitting procedure we performed an additional quality test: a set of 50 randomized \textit{ab initio} points were calculated at the CCSD(T)-F12/aVDZ level of theory in the attractive region of the PES. The final accuracy of the fit was better than 1 cm$^{-1}$ for the attractive part of the PES. The long-range region of the potential ($R\ge19$ a$_0$) was extrapolated using an inverse exponent expansion implemented in the {\small MOLSCAT} code \cite{molscat95}.

\subsection{Description of the potential energy surface}
\begin{figure*}[]
    \centering
    \includegraphics[width=0.45\linewidth]{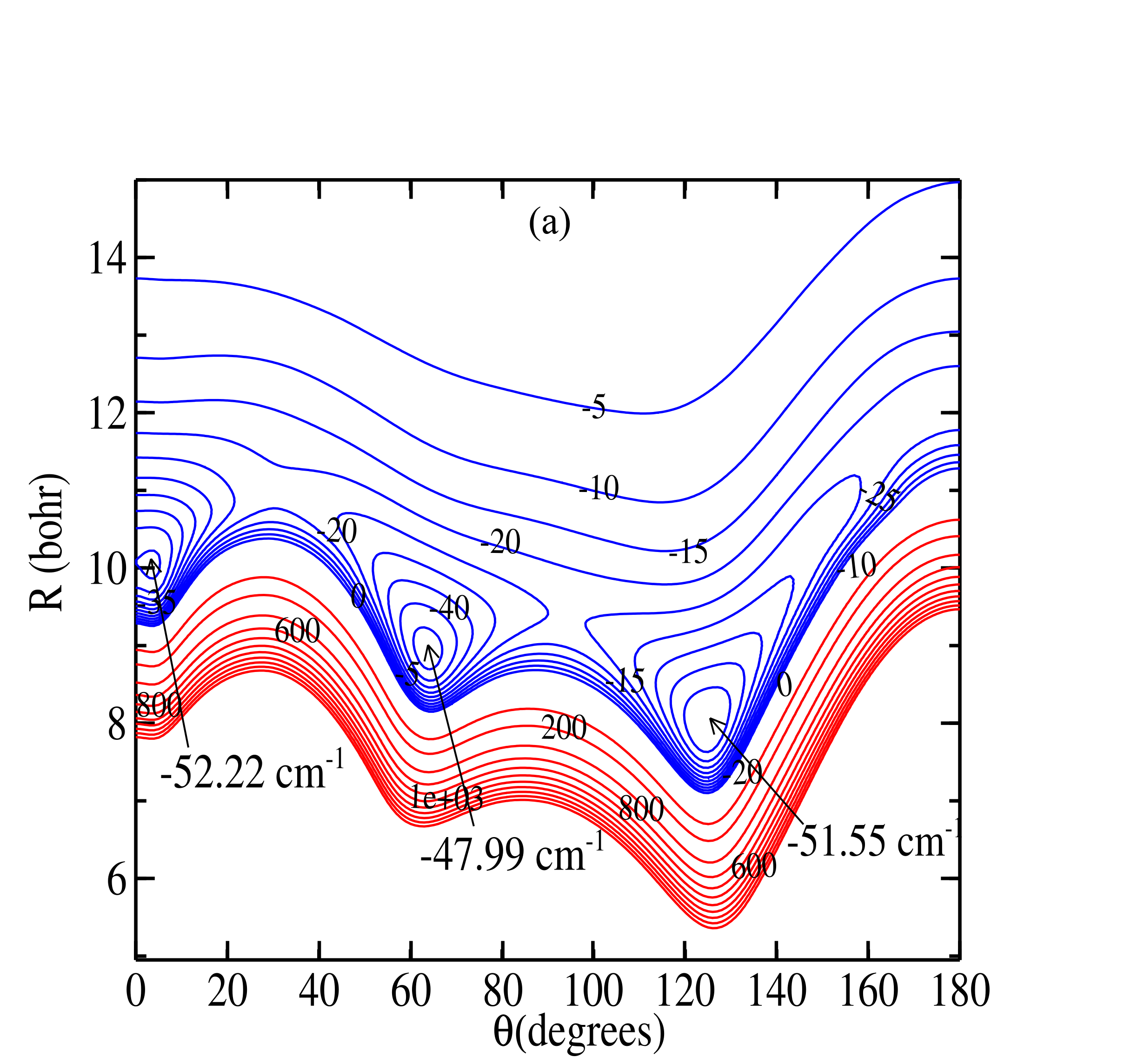}
    \includegraphics[width=0.45\linewidth]{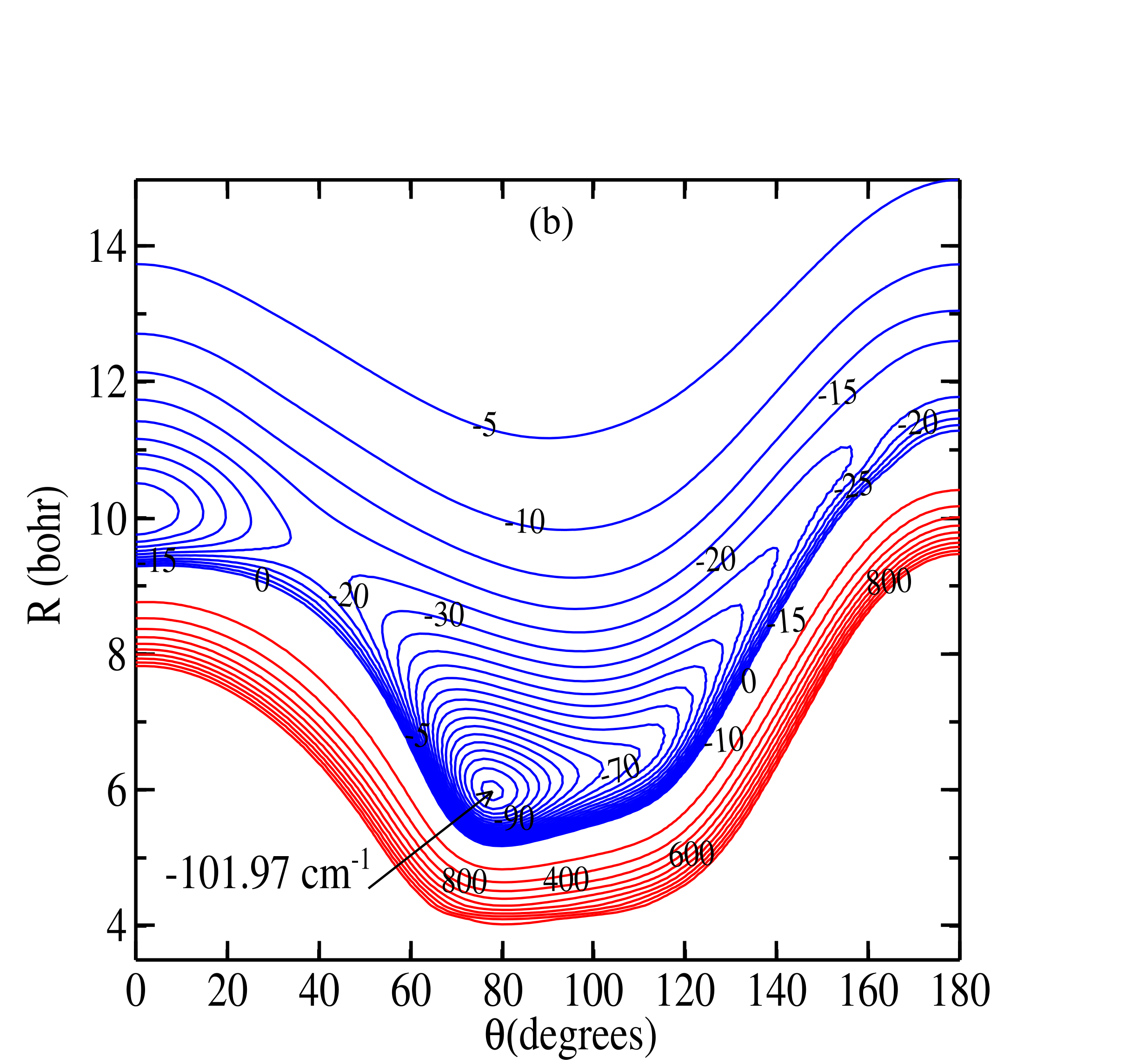}
    \includegraphics[width=0.45\linewidth]{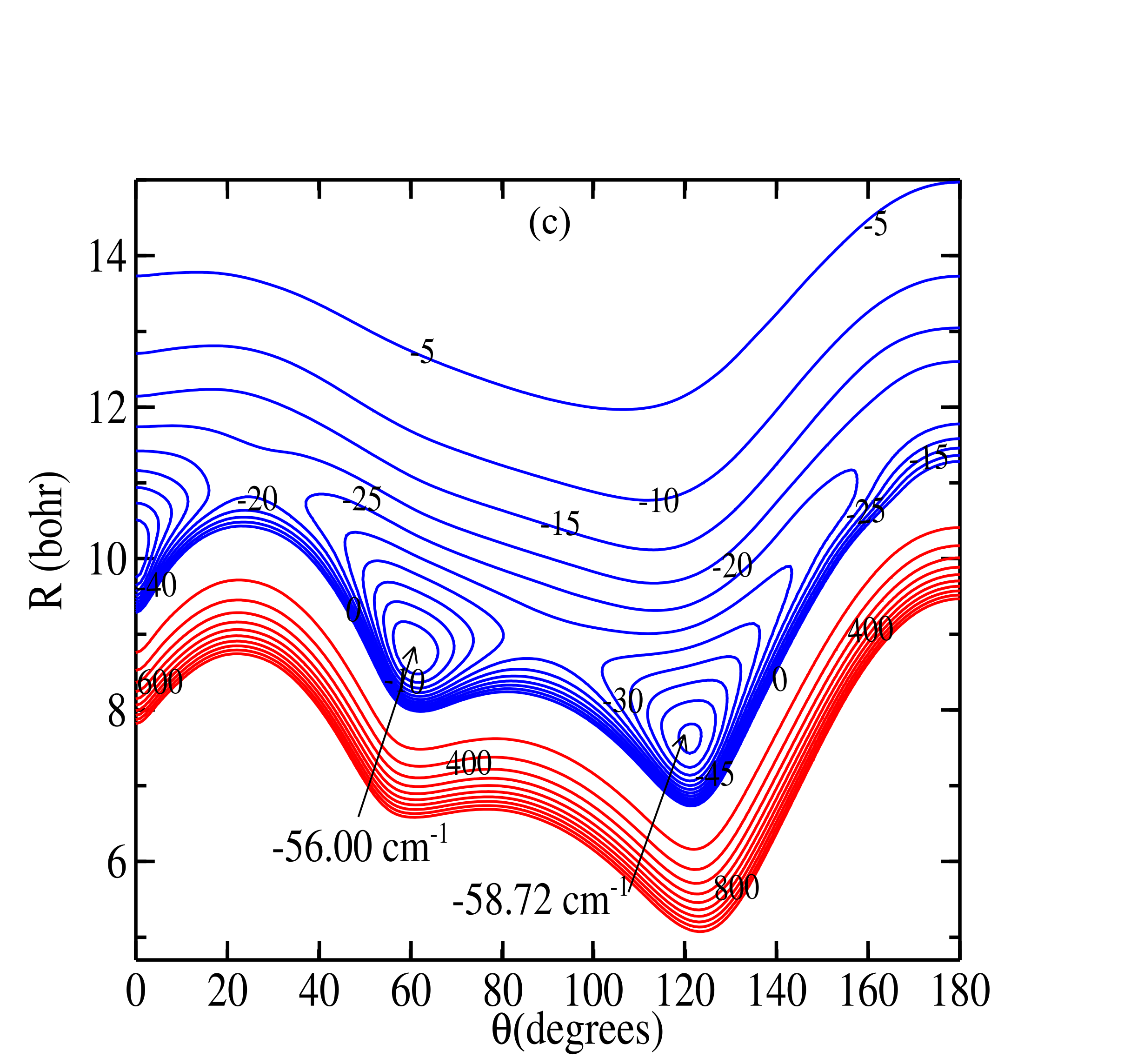}
    \includegraphics[width=0.45\linewidth]{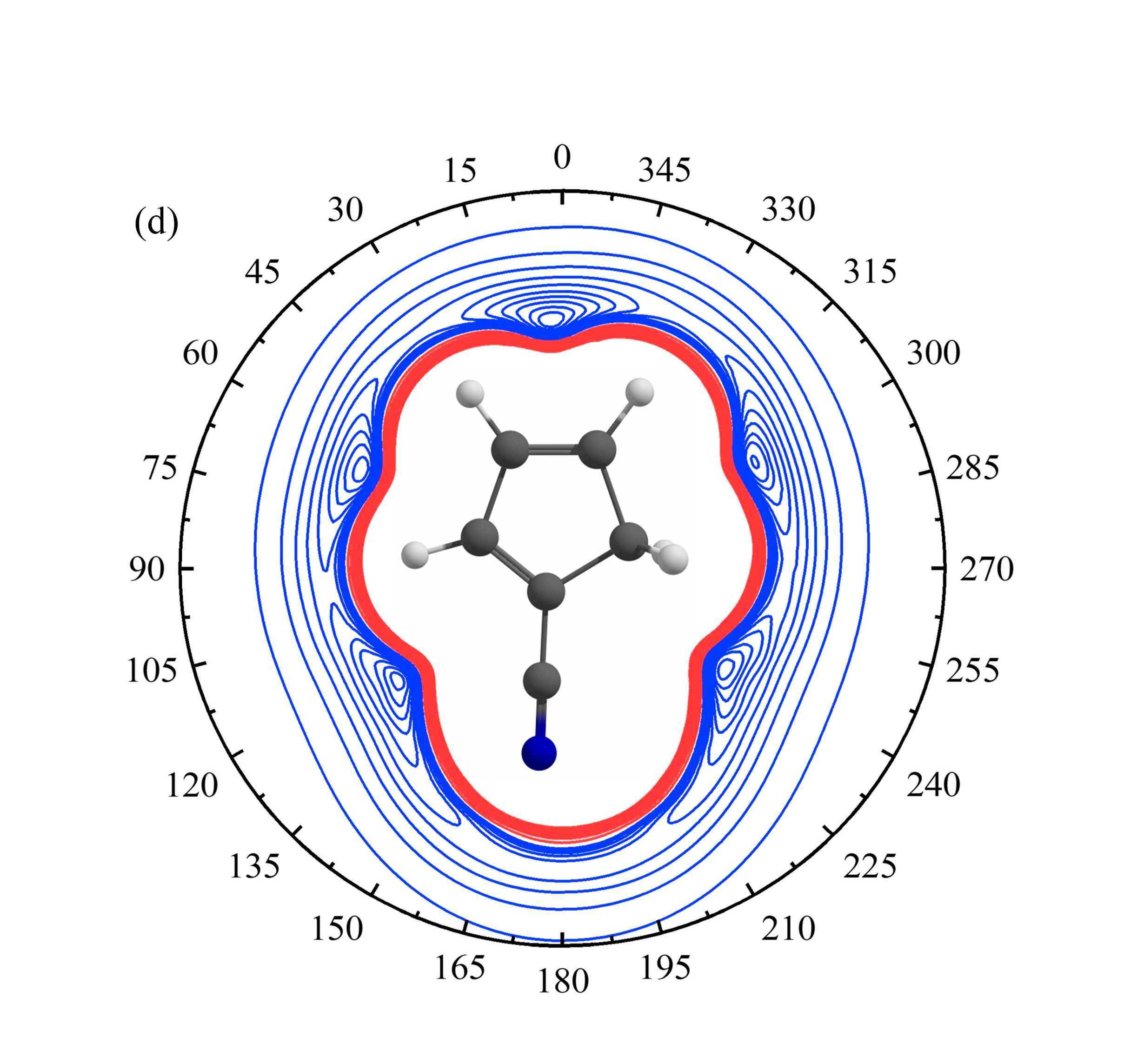}
    \caption{Contour plots of the interaction potential of the 1-cyano-CPD--He complex (blue - attractive part of the PES, red - repulsive part of the PES, energy given in cm$^{-1}$). Panels (a), (b), and (c) depict the PES as a function of the two Jacobi coordinates $R$ and $\theta$ for $\phi=0^{\circ}$, $\phi=95^{\circ}$, $\phi=180^{\circ}$, respectively. Panel (d) shows a two-dimensional contour plot in polar coordinates where the He atom movement is confined within the molecular plane.}
    \label{fig:CPD-PLOTS}
\end{figure*}

Figure \ref{fig:CPD-PLOTS} illustrates two-dimensional contour plots of the interaction potential of the 1-cyano-CPD -- He complex. Panels (a) and (c) represent the rotation of the He atom in the molecular plane, panel (b) represents the movement of the helium atom out of the molecular plane. Panel (d) depicts the rotation of the He atom in the molecular plane in polar coordinates.\\
The global minimum of the PES corresponds to the He atom hovering above the molecular plane close to the sp$^3$ carbon atom. The geometry of the minimum is $\phi=95^{\circ}$, $\theta=77^{\circ}$, $R=6.1$ with a well depth of $V=-101.8$cm$^{-1}$, as shown in panel (b).\\
Several local minima are observed for the rotation of He atom in the molecular plane. For $\phi=0^{\circ}$ (panel (a)) we locate three minima at $\theta=3^{\circ}$ and $R=10.0$ a$_0$, $\theta=63^{\circ}$ and $R=8.9$ a$_0$, $\theta=125^{\circ}$ and $R=7.9$a$_0$, with well depths of -52.2, -48.0 and -51.6 cm$^{-1}$ respectively. The first and second wells are separated by a barrier of 31 cm$^{-1}$, whereas a 27 cm$^{-1}$ barrier lies between the second and the third wells. The location of the first minimum at $\theta=3^{\circ}$ instead of expected $\theta=0^{\circ}$ is related to the definition of the coordinate system. Considering the asymmetric geometry of 1-cyano-CPD, we observe a slight shift of the mass center and inertia axes, thus the coordinate axis \textit{z} is not aligned with the C$\equiv$N bond. For $\phi=180^{\circ}$ (panel (c)) the potential wells are found at $\theta=61^{\circ}$ and $R=8.8$ a$_0$, $\theta=122^{\circ}$ and $R=7.6$ a$_0$, with depths of -56.0 and -58.7 cm$^{-1}$ separated by a barrier of 26 cm$^{-1}$. Thus, for the rotation in the molecular plane we report 5 minima which can be clearly seen in panel (d). These minima correspond to helium approaching between two hydrogen atoms or between a hydrogen atom and the nitrile group. It is worth noting that the presence of an additional H atom (on the sp$^3$ carbon) leads to deeper minima (-56.0 and -58.72 vs. -47.99 and -51.6 cm$^{-1}$) as shown in panels (a) and (c).\\
Overall, the features observed for the PES of 1-cyano-CPD--He system are consistent with those found in related systems such as benzonitrile \cite{ben2024rotational} and cyclopentadiene \cite{demes2024first}: for the in-plane movement of the helium atom the number of minima corresponds to the number of C-C bonds within the cycle (5 for 1-cyano-CPD and cyclopentadiene, 6 for benzonitrile). The minima located around sp$^2$ carbon atoms also demonstrate a similar well depth of about -50 cm$^{-1}$. Moreover, these two systems have an additional plane of symmetry, so the location of their respective global minima at $\phi=90^{\circ}$ is aligned with the mirror plane and only reflects the symmetric nature of the molecule. However, for cyclopentadiene, we also observe that the global minimum is shifted towards the sp$^3$ carbon as in the case of 1-cyano-CPD, whereas for benzonitrile the location of the global minimum is shifted towards the nitrile group. 
\section{Dynamics}\label{dynamics}
\subsection{Rotational Hamiltonian}
\begin{figure}[]
    \centering
    \includegraphics[width=0.95\linewidth]{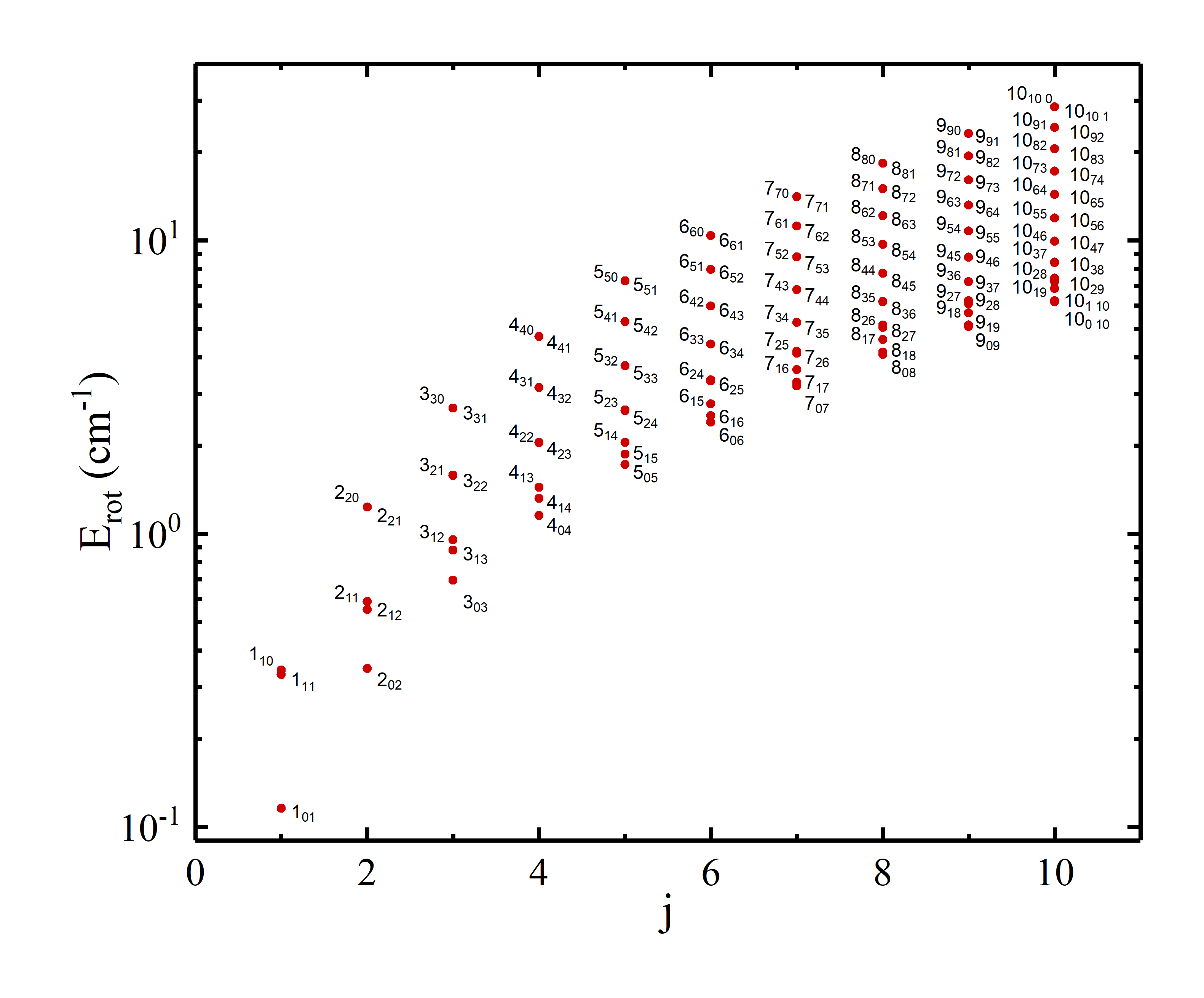}
    \caption{Rotational levels of 1-cyano-CPD up to $j = 10$. The $0_{00}$ level is not shown, due to having the energy of 0 cm$^{-1}$}
    \label{fig:rotlvl}
\end{figure}

1-cyano-CPD is a near prolate asymmetric top molecule with a rotational Hamiltonian written as \cite{schmitt2018structures}:
\begin{equation}\label{Hrot}
    H_{rot}=Aj_x^2+Bj_y^2+Cj_z^2-D_jj^4-D_{jk}j^2j_z^2-D_kj_z^4
\end{equation}
where $A$, $B$ and $C$ are rotational constants, $D_j$, $D_{jk}$, $D_k$ are first order centrifugal distortion constants of 1-cyano-CPD and $j_x$, $j_y$, $j_z$ are the projections of the angular momentum \textbf{j} along the principal inertia axes: $j^2=j^2_x+j^2_y+j^2_z$. The rotational levels of asymmetric top molecules are labelled by $k_a$ and $k_c$ numbers representing the projections of the quantum number $k$ along the axes of symmetry in the prolate and oblate symmetric top limits. The relation between $k_a$ and $k_c$ can also be expressed through $\tau$: $\tau$ = $k_a - k_c$. The rotational constants $A=0.279$, $B=0.064$, $C=0.052$ and centrifugal distortion constants $D_j = 2.669 \times 10^{-9}$, $D_{jk} = 8.108 \times 10^{-8}$ and $D_k = 5.859 \times 10^{-8}$ (all in cm$^{-1}$) were taken from the literature \cite{sakaizumi1987microwave,mccarthy2021interstellar}. The rotational energy level diagram of 1-cyano-CPD is shown in Fig. \ref{fig:rotlvl}. \\
The eigenfunctions $\vert j\tau m \rangle $ of the rotational Hamiltonian in equation \ref{Hrot} are expressed as linear combinations of symmetric top wavefunctions $\vert jkm \rangle $:
\begin{equation}\label{wavef}
    \vert j\tau m \rangle = \sum_k a_{\tau k}\vert jkm\rangle
\end{equation}
where $k$ denotes the projection of $j$ along the $z$-axis of the body-fixed frame, $m$ is the projection on the space-fixed $Z$-axis and $\tau$ is an integer $-j\leq\tau\leq j$ which orders the energy levels for a given value of $j$ \cite{flower2007molecular}.\\
\subsection{Scattering calculations}\label{scattering}
\begin{figure*}[]
    \centering
    \includegraphics[width=0.49\linewidth]{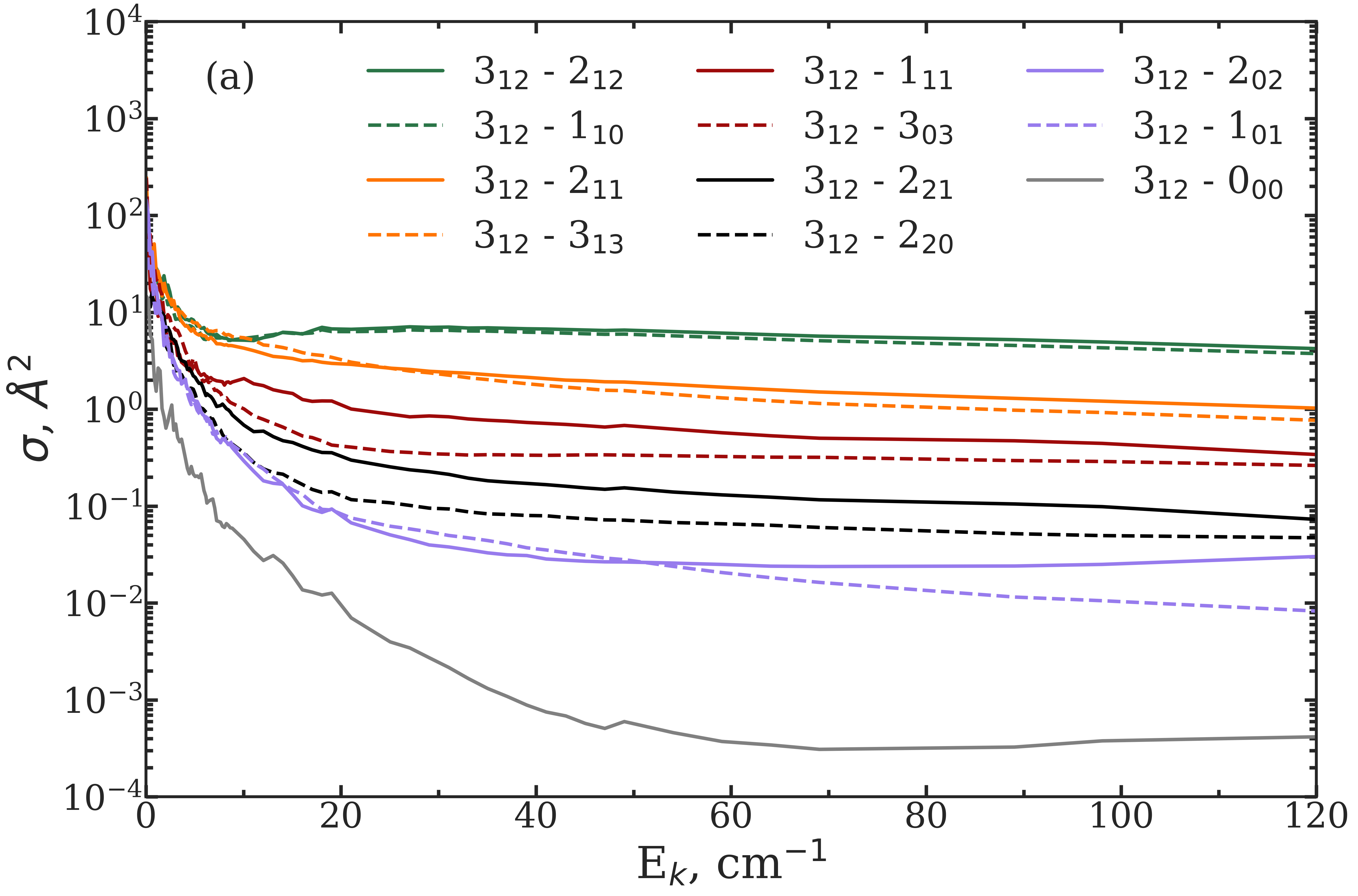}
    \includegraphics[width=0.49\linewidth]{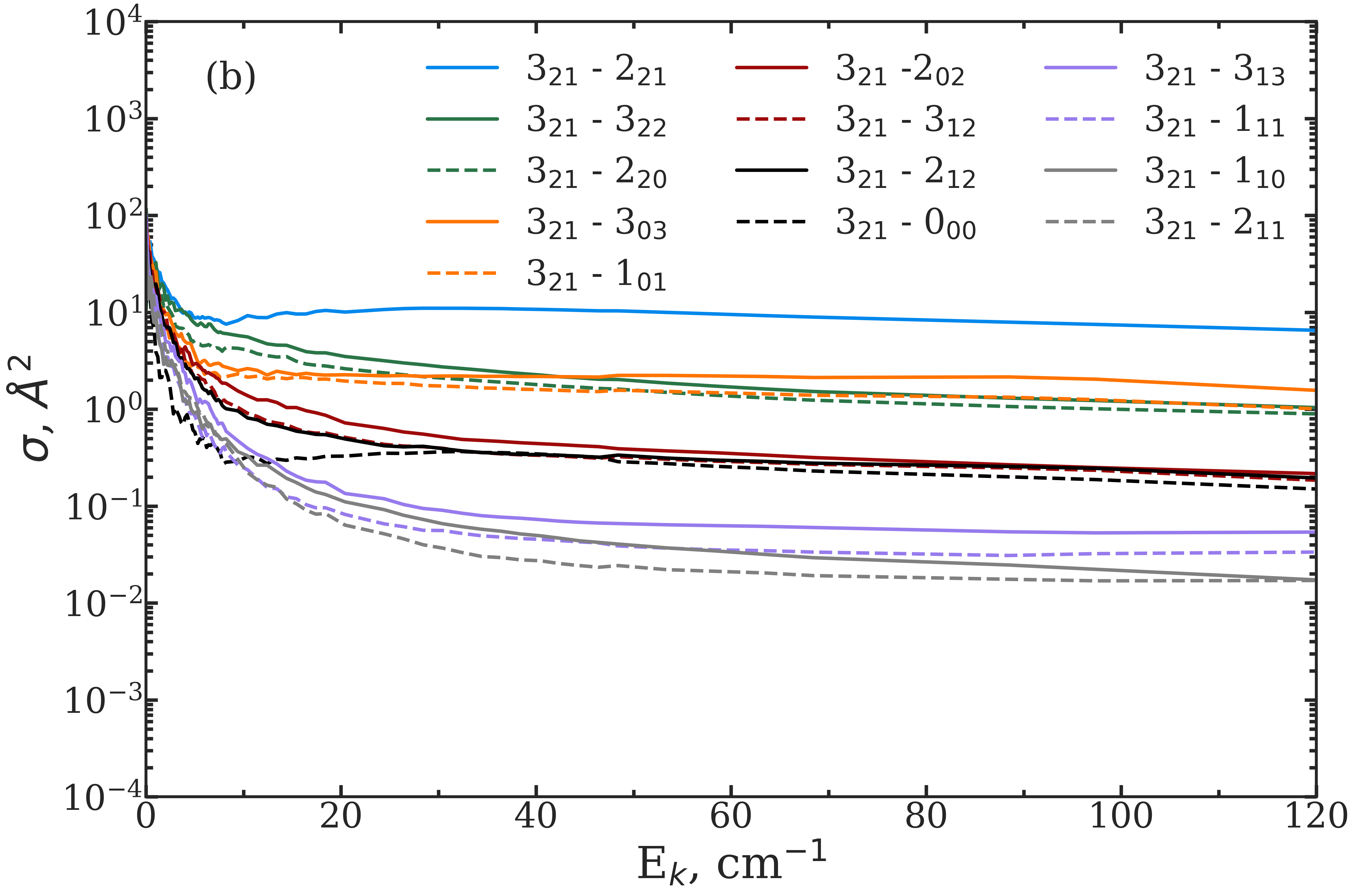}
    \caption{Kinetic energy dependence of the rotational de-excitation cross sections $\sigma_{jk_ak_c \xrightarrow{} j'k'_ak'_c}$ of 1-cyano-CPD by He atoms from the $3_{12}$ (panel a) and $3_{21}$ (panel b) initial states.}
    \label{fig:from3_}
\end{figure*}

\begin{figure*}[]
    \centering
    \includegraphics[width=0.49\linewidth]{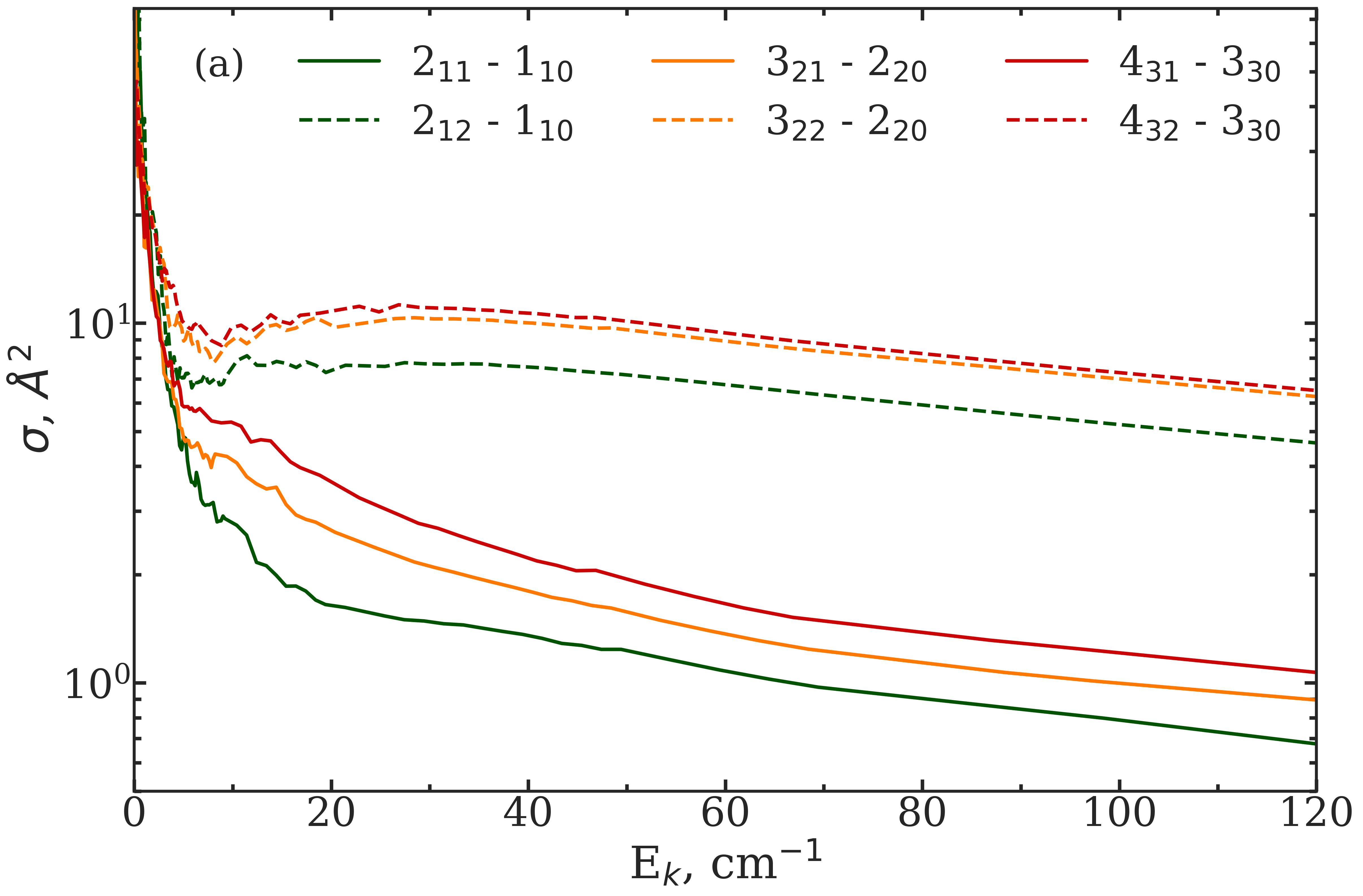}
    \includegraphics[width=0.49\linewidth]{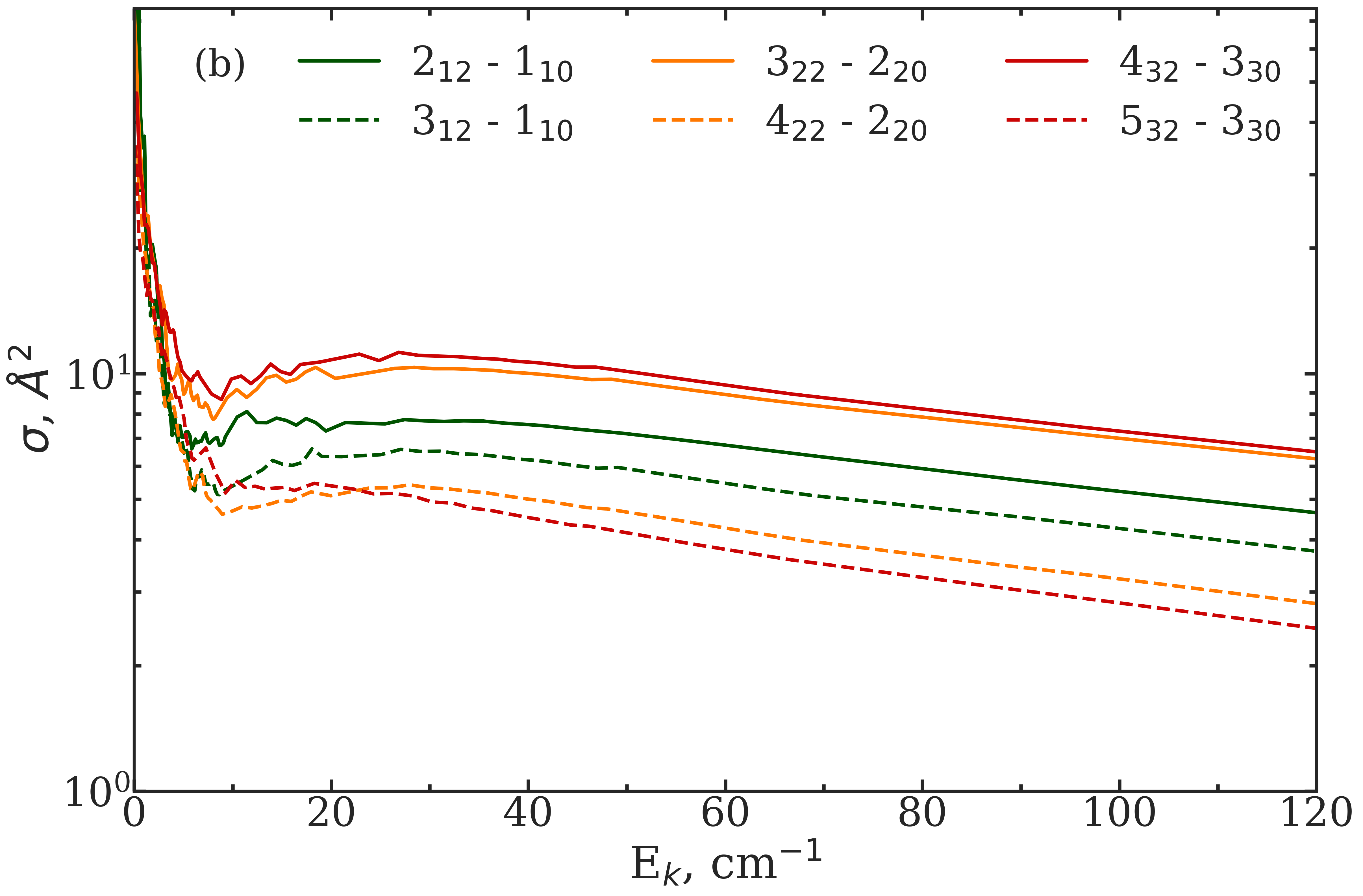}
    \caption{Kinetic energy dependence of the rotational de-excitation cross sections $\sigma_{jk_ak_c \xrightarrow{} j'k'_ak'_c}$ of 1-cyano-CPD by collisions with He atoms for $\Delta j=1$, $\Delta k_a=0$, and $\Delta k_c=1$ or 2 (panel a), $\Delta j=1$ or 2, $\Delta k_a=0$, and $\Delta k_c=2$ (panel b).}
    \label{fig:prop}
\end{figure*}

To our current knowledge, 1-cyano-CPD has only been observed in TMC-1 \cite{mccarthy2021interstellar}. Considering the physical conditions of this dense dark molecular cloud, the emphasis of this study is the collisional excitation in the low-energy regime. Here we aim to provide collisional cross sections for $E_{\textrm{tot}}=E_{\textrm{k}}+E_{\textrm{rot}} \leq 125$ cm$^{-1}$, where $E_\textrm{k}$ is the kinetic energy of the collision and $E_{\textrm{rot}}$ is the energy of the initial rotational level. Transitions up to $j = 9$ were observed in the ISM, therefore, we focus on obtaining converged cross sections for transitions involving levels up to $j = 10$, taking into consideration a sufficient number of open energy channels.\\
The rotational cross sections $\sigma_{jk_ak_c \xrightarrow{} j'k'_ak'_c}$ were computed using the quantum close-coupling (CC) method \cite{arthurs1960theory} for collisions of asymmetric top molecules with helium atoms \cite{green1976rotational,garrison1976coupled} implemented in a parallelized version of {\small MOLSCAT} code \cite{molscat95}. Radial coefficients $v_{lm}(R)$ obtained in section \ref{fit-subsect} were employed in the scattering code. The coupled equations were solved with the diabatic log-derivative propagator \cite{manolopoulos1986improved}. The reduced mass of the interacting system was taken as $\mu = 3.81156$ a.u. (isotopes $^{12}$C, $^{14}$N, $^{1}$H and $^4$He). Several cross section convergence tests were performed to set the integration boundaries of the propagator: $R_{\min}$ and $R_{\max}$ were fixed at 4.0 a$_0$ and  40 a$_0$, respectively. Considering the dense structure of rotational levels, an energy cutoff defined by the parameter $E_{\max}$ was employed to improve the calculation runtimes. We tested the convergence with respect to the size of the rotational basis set (defined by parameter $j_{\max}$) in conjunction with $E_{\max}$ for several total energies. The resulting parameters were defined as follows. For energies $E_{\textrm{tot}} \leq 50$ cm$^{-1}$ the parameters were set as $j_{\max} = 18$ and $E_{\max} = 50$ cm$^{-1}$, for $50 < E_{\textrm{tot}} \leq 100$ cm$^{-1}$ as $j_{\max} = 19$ and $E_{\max} = 100$ cm$^{-1}$, and for $100 < E_{\textrm{tot}} \leq 125$ as $j_{\max} = 22$ and $E_{\max} = 125$ cm$^{-1}$. The values were selected so that the inelastic cross sections were converged within 5\%.  
The cross sections were calculated over a grid of 129 energy points with a variable step size to ensure a correct description of the resonance region for the various initial rotational levels. The maximum value of the total angular momentum $J$ was chosen so that the inelastic cross sections were converged within 0.05 \AA$^2$.\\
Figure \ref{fig:from3_} displays rotational quenching cross sections for the initial rotational levels $j_{k_a k_c}=3_{12}$ and $j_{k_a k_c}=3_{21}$ with corresponding energies $E_{\textrm{rot}}=0.95$ and 1.58 cm$^{-1}$ as seen in Figure \ref{fig:rotlvl}. At low energies ($E_k \leq$ 20 cm$^{-1}$) the cross sections demonstrate similar features of a dense resonant structure. This can be attributed to the formation of a quasi-bound collisional complex caused by the presence of the attractive potential well of 101.8 cm$^{-1}$. \\
For kinetic energies above 20 cm$^{-1}$ we observe propensity rules favoring transitions with $\Delta k_a=0$: 3$_{12}$\,--\,2$_{12}$, 3$_{12}$\,--\,1$_{10}$, 3$_{12}$\,--\,2$_{11}$, 3$_{12}$\,--\,3$_{13}$, 3$_{21}$\,--\,2$_{21}$, 3$_{21}$\,--\,3$_{22}$, 3$_{21}$\,--\,2$_{20}$.
Among those, the largest cross sections are systematically observed for transitions with $\Delta j=1$, $\Delta k_c=0$ (3$_{12}$\,--\,2$_{12}$ and 3$_{21}$\,--\,2$_{21}$) and $\Delta j=2$, $\Delta k_c=2$ (3$_{12}$\,--\,1$_{10}$). Such a transition from the level $3_{21}$ is impossible, since the corresponding final level doesn't exist, hence why it is not presented in panel (b). Moreover, the cross sections of both aforementioned transition types are almost identical over the considered energy range, as can be seen in panel (a). Panel (b) also demonstrates that the dominant transitions with $\Delta k_a=0$ are followed by those with $\Delta k_a=2$ (3$_{21}$\,--\,3$_{03}$, 3$_{21}$\,--\,1$_{01}$, 3$_{21}$\,--\,3$_{03}$).\\
In Figure \ref{fig:prop} we illustrate more state-to-state rotational quenching cross sections to further analyze the propensity rules within the dominant $\Delta k_a=0$ transitions. Here we report the cross sections for collisions of 1-cyano-CPD with He atoms as a function of kinetic energy for transitions with specific values of $\Delta j$, $\Delta k_a$ and $\Delta k_c$. Panel (a) demonstrates a strong propensity favoring $\Delta k_c=2$ transitions within the $\Delta k_a=0$ ladder. This confirms that the propensity rules observed for the transitions from the initial level $3_{12}$ in Figure \ref{fig:from3_} also hold for other initial levels where $\Delta k_a=0$, $\Delta k_c=2$ de-excitation is possible. Panel (b) allows us to look into the $\Delta j$ propensity, which is not as prominent neither in $\Delta k_a=0$ ladder nor in other ladders.\\
Thus, the trend is as follows: a clear propensity rule favors $\Delta k_a=0$ transitions, with $\Delta k_c=0$ and 2 being the most dominant among those. Transitions of this type imply the conservation of angular momentum along the largest inertia axis \textit{a}. Such propensity has already been observed for other asymmetric top COMs: propylene oxide \cite{faure2019interaction}, benzonitrile \cite{ben2024rotational}, and vinyl cyanide \cite{sogomonyan2025rotational}. 
\subsection{Rate coefficients}\label{ratecoefficients}
\begin{figure}[]
    \centering
    \includegraphics[width=0.95\linewidth]{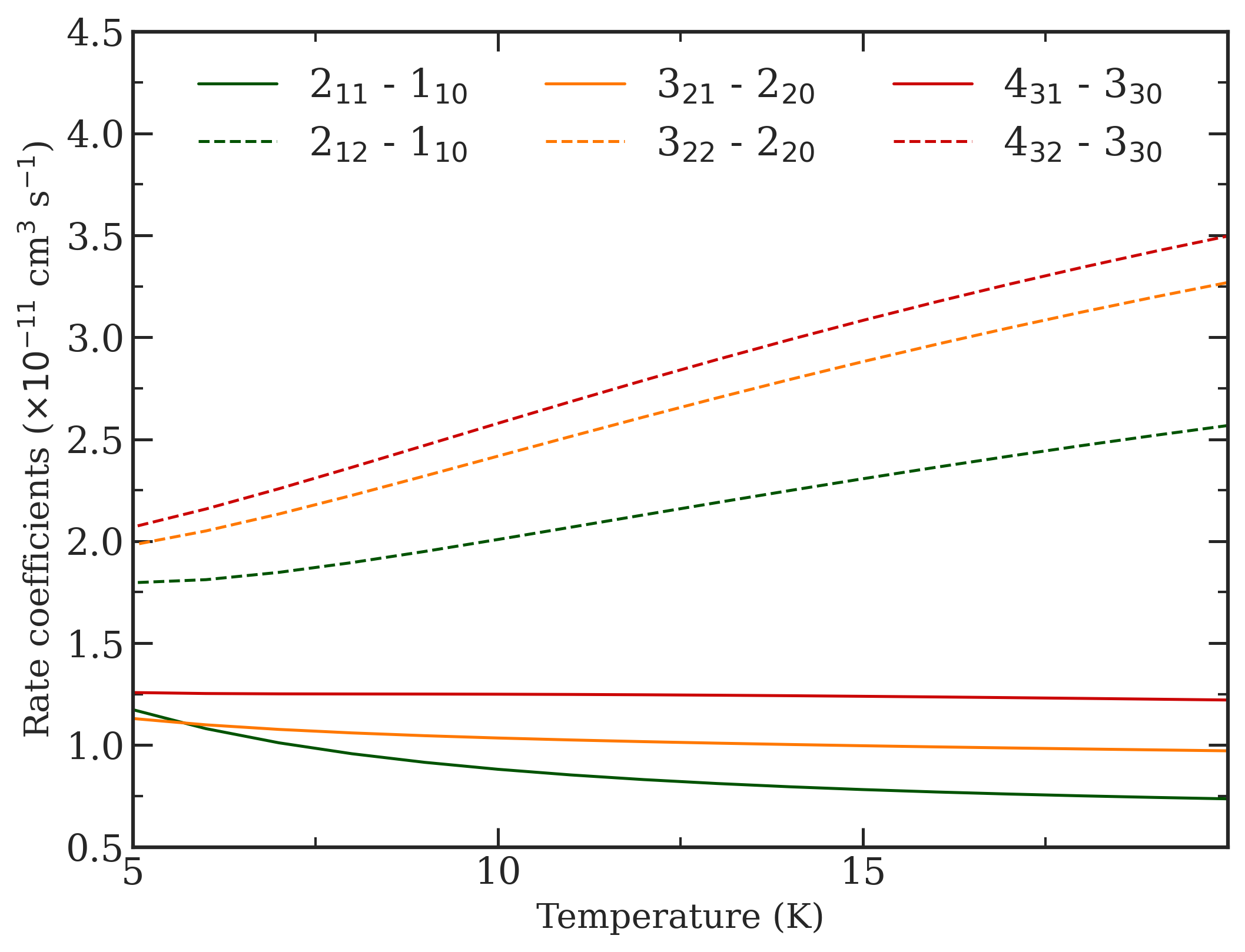}
    \caption{Temperature dependence of the rotational rate coefficients for transitions $jk_ak_c \rightarrow j'k'_ak'_c$ of 1-cyano-CPD in collision with He atoms for $\Delta j=1$, $\Delta k_a=0$, and $\Delta k_c=1$ and 2.}
    \label{fig:ratecoef}
\end{figure}

\begin{figure}[]
    \centering
    \includegraphics[width=0.99\linewidth]{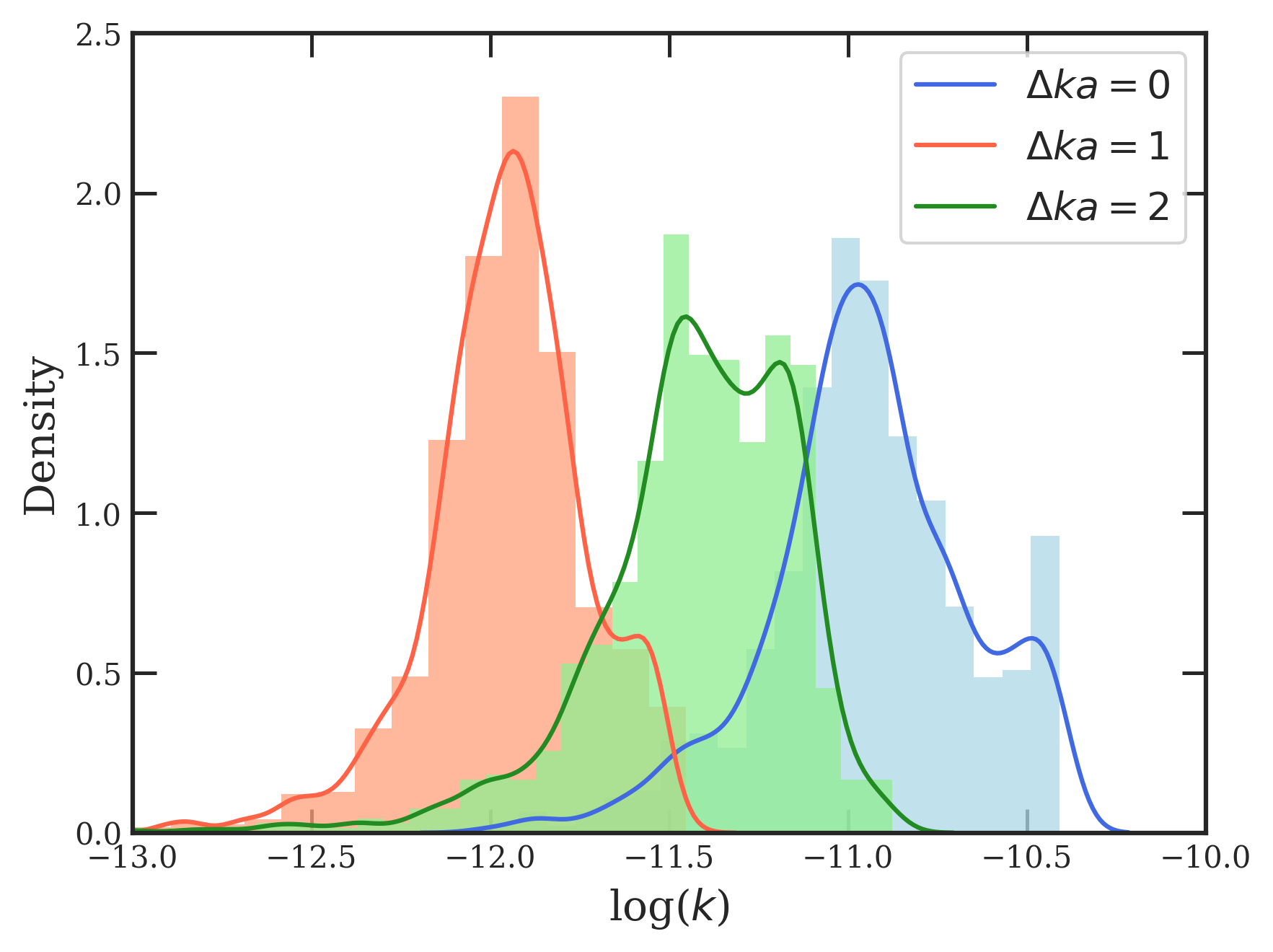}
    \caption{Histogram distribution of the logarithm base 10 of the rotational quenching rate coefficients $jk_ak_c \rightarrow j'k'_ak'_c$ at $T=20$ K for $\Delta k_a=0, 1$ and 2 for 4753 transitions with $j$ and $j'\leq9$. The lines show a kernel density estimate with bandwidth of 0.8.}
    \label{fig:hist}
\end{figure}

From state-to-state cross sections calculated for total energies up to 125 cm$^{-1}$ we computed the corresponding rate coefficients for transitions involving the lowest rotational levels for kinetic temperatures between 5 and 20 K by averaging the cross sections over a Maxwell-Boltzmann distribution:
\begin{equation}
 k_{i \rightarrow f}(T)=\biggl(\frac{8}{\pi\mu\beta}\biggl)^{\frac{1}{2}}\beta^2\int_0^{\infty} E_k \sigma_{i \rightarrow f}(E_k)e^{-\beta E_k} dE_k
\end{equation}
where $\beta$=$1/k_BT$ and $k_B$, $T$ and $\mu$ are the Boltzmann constant, the kinetic temperature and the collision reduced mass respectively.\\
Rotational quenching rate coefficients for 1-cyano-CPD--He system corresponding to $\Delta j=1$, $\Delta k_a=0$, $\Delta k_c=1$ and 2 transitions are presented in Figure \ref{fig:ratecoef}. A more noticeable temperature dependence is observed for the rate coefficients corresponding to $\Delta k_c=2$ transitions, whereas the decrease in the $\Delta k_c=1$ rate coefficients is only minor. Additionally, the magnitude of the rate coefficients associated with $\Delta k_c=2$ transitions is larger than that of $\Delta k_c=1$ transitions, which reflects the behavior of the corresponding cross sections.\\
Figure \ref{fig:hist} illustrates the distribution of 4753 rotational quenching rate coefficients for all initial and final states with $j\leq9$ separated by the value of $\Delta k_a$. This distribution exhibits the clear $\Delta k_a=0$ propensity observed for the cross sections discussed in section \ref{scattering}. It is also clear that the next closest rate coefficients are those within the $\Delta k_a=2$ ladder, which reflects the trend demonstrated by the corresponding cross sections. As discussed in section \ref{scattering} for inelastic cross sections, propensity rules in $\Delta j$ or $\Delta k_c$ are much less prominent.\\
\section{Conclusions} \label{conclusions}
In this work we have presented the first potential energy surface for the 1-cyano-CPD--He system. The three-dimensional PES was computed at the CCSD(T)-F12/aVDZ level of theory. The global minimum of -101.8 cm$^{-1}$ was found at $\phi=95^{\circ}$, $\theta=77^{\circ}$, $R=6.1$ a$_0$, corresponding to the He atom hovering above the molecular plane. The interaction potential exhibits strong anisotropy with respect to the angle $\theta$, leading to five distinct local minima separated by energy barriers associated with the rotation of the He atom within the molecular plane.\\
The PES was subsequently employed in quantum scattering calculations to compute state-to-state rotational quenching cross sections using the quantum-mechanical close coupling method. Rotational cross sections were calculated for total energies reaching up to 125 cm$^{-1}$. By thermally averaging the cross sections over the Maxwell-Boltzmann distribution, we obtained the state-to-state collisional rate coefficients for temperatures up to 20 K. Propensity rules favoring transitions with $\Delta k_a = 0$ were observed, with additional trends confirming the systematic dominance of transitions with $\Delta k_a = 0$, $\Delta k_c = 0$ and $\Delta k_a = 0$, $\Delta k_c = 2$.\\
Future work will examine wether the LTE conditions are fulfilled for 1-cyano-CPD by means of radiative transfer calculations. Non-LTE modeling of the molecular lines will further the understanding of the non-LTE effects for cyclic COMs and allow to better constrain the molecular abundance.\\
In this work we have focused on the most stable isomer of cyanocyclopentadiene, however, a study of the other detected isomer (2-cyano-CPD) could also be of interest. Although the interaction potential of the two isomers with He should be roughly similar, the collisional excitation rate coefficients of COMs can still display significant isomerism effects \cite{ben2022interaction,ben2023collisional}.

\begin{acknowledgments}
JL acknowledges support from KU Leuven through Project no. C14/22/082. MBK acknowledges support from the Research Foundation-Flanders (FWO). The scattering calculations presented in this work were performed on the VSC clusters (Flemish Supercomputer Center), funded by the Research Foundation-Flanders (FWO) and the Flemish Government.
\end{acknowledgments}

\section{Data Availability Statement}
Fortran subroutine for potential energy surface calculations from radial coefficients is available as supporting information. Collisional rate coefficients will be made available in the BASECOL \cite{dubernet2024basecol2023} database: \url{https://basecol.vamdc.eu/}.

\bibliography{aipsamp}

\begin{thebibliography}{40}%
\makeatletter
\providecommand \@ifxundefined [1]{%
 \@ifx{#1\undefined}
}%
\providecommand \@ifnum [1]{%
 \ifnum #1\expandafter \@firstoftwo
 \else \expandafter \@secondoftwo
 \fi
}%
\providecommand \@ifx [1]{%
 \ifx #1\expandafter \@firstoftwo
 \else \expandafter \@secondoftwo
 \fi
}%
\providecommand \natexlab [1]{#1}%
\providecommand \enquote  [1]{``#1''}%
\providecommand \bibnamefont  [1]{#1}%
\providecommand \bibfnamefont [1]{#1}%
\providecommand \citenamefont [1]{#1}%
\providecommand \href@noop [0]{\@secondoftwo}%
\providecommand \href [0]{\begingroup \@sanitize@url \@href}%
\providecommand \@href[1]{\@@startlink{#1}\@@href}%
\providecommand \@@href[1]{\endgroup#1\@@endlink}%
\providecommand \@sanitize@url [0]{\catcode `\\12\catcode `\$12\catcode `\&12\catcode `\#12\catcode `\^12\catcode `\_12\catcode `\%12\relax}%
\providecommand \@@startlink[1]{}%
\providecommand \@@endlink[0]{}%
\providecommand \url  [0]{\begingroup\@sanitize@url \@url }%
\providecommand \@url [1]{\endgroup\@href {#1}{\urlprefix }}%
\providecommand \urlprefix  [0]{URL }%
\providecommand \Eprint [0]{\href }%
\providecommand \doibase [0]{http://dx.doi.org/}%
\providecommand \selectlanguage [0]{\@gobble}%
\providecommand \bibinfo  [0]{\@secondoftwo}%
\providecommand \bibfield  [0]{\@secondoftwo}%
\providecommand \translation [1]{[#1]}%
\providecommand \BibitemOpen [0]{}%
\providecommand \bibitemStop [0]{}%
\providecommand \bibitemNoStop [0]{.\EOS\space}%
\providecommand \EOS [0]{\spacefactor3000\relax}%
\providecommand \BibitemShut  [1]{\csname bibitem#1\endcsname}%
\let\auto@bib@innerbib\@empty
\bibitem [{\citenamefont {Yamamoto}(2017)}]{yamamoto2017introduction}%
  \BibitemOpen
  \bibfield  {author} {\bibinfo {author} {\bibfnamefont {S.}~\bibnamefont {Yamamoto}},\ }\href@noop {} {\bibfield  {journal} {\bibinfo  {journal} {Editorial: Springer}\ }\textbf {\bibinfo {volume} {614}} (\bibinfo {year} {2017})}\BibitemShut {NoStop}%
\bibitem [{\citenamefont {Tielens}(2008)}]{tielens2008interstellar}%
  \BibitemOpen
  \bibfield  {author} {\bibinfo {author} {\bibfnamefont {A.~G.}\ \bibnamefont {Tielens}},\ }\href@noop {} {\bibfield  {journal} {\bibinfo  {journal} {Annu. Rev. Astron. Astrophys.}\ }\textbf {\bibinfo {volume} {46}},\ \bibinfo {pages} {289} (\bibinfo {year} {2008})}\BibitemShut {NoStop}%
\bibitem [{\citenamefont {Wenzel}\ \emph {et~al.}(2024)\citenamefont {Wenzel}, \citenamefont {Cooke}, \citenamefont {Changala}, \citenamefont {Bergin}, \citenamefont {Zhang}, \citenamefont {Burkhardt}, \citenamefont {Byrne}, \citenamefont {Charnley}, \citenamefont {Cordiner}, \citenamefont {Duffy} \emph {et~al.}}]{wenzel2024detection}%
  \BibitemOpen
  \bibfield  {author} {\bibinfo {author} {\bibfnamefont {G.}~\bibnamefont {Wenzel}}, \bibinfo {author} {\bibfnamefont {I.~R.}\ \bibnamefont {Cooke}}, \bibinfo {author} {\bibfnamefont {P.~B.}\ \bibnamefont {Changala}}, \bibinfo {author} {\bibfnamefont {E.~A.}\ \bibnamefont {Bergin}}, \bibinfo {author} {\bibfnamefont {S.}~\bibnamefont {Zhang}}, \bibinfo {author} {\bibfnamefont {A.~M.}\ \bibnamefont {Burkhardt}}, \bibinfo {author} {\bibfnamefont {A.~N.}\ \bibnamefont {Byrne}}, \bibinfo {author} {\bibfnamefont {S.~B.}\ \bibnamefont {Charnley}}, \bibinfo {author} {\bibfnamefont {M.~A.}\ \bibnamefont {Cordiner}}, \bibinfo {author} {\bibfnamefont {M.}~\bibnamefont {Duffy}},  \emph {et~al.},\ }\href@noop {} {\bibfield  {journal} {\bibinfo  {journal} {Science}\ }\textbf {\bibinfo {volume} {386}},\ \bibinfo {pages} {810} (\bibinfo {year} {2024})}\BibitemShut {NoStop}%
\bibitem [{\citenamefont {McGuire}\ \emph {et~al.}(2021)\citenamefont {McGuire}, \citenamefont {Loomis}, \citenamefont {Burkhardt}, \citenamefont {Lee}, \citenamefont {Shingledecker}, \citenamefont {Charnley}, \citenamefont {Cooke}, \citenamefont {Cordiner}, \citenamefont {Herbst}, \citenamefont {Kalenskii} \emph {et~al.}}]{mcguire2021detection}%
  \BibitemOpen
  \bibfield  {author} {\bibinfo {author} {\bibfnamefont {B.~A.}\ \bibnamefont {McGuire}}, \bibinfo {author} {\bibfnamefont {R.~A.}\ \bibnamefont {Loomis}}, \bibinfo {author} {\bibfnamefont {A.~M.}\ \bibnamefont {Burkhardt}}, \bibinfo {author} {\bibfnamefont {K.~L.~K.}\ \bibnamefont {Lee}}, \bibinfo {author} {\bibfnamefont {C.~N.}\ \bibnamefont {Shingledecker}}, \bibinfo {author} {\bibfnamefont {S.~B.}\ \bibnamefont {Charnley}}, \bibinfo {author} {\bibfnamefont {I.~R.}\ \bibnamefont {Cooke}}, \bibinfo {author} {\bibfnamefont {M.~A.}\ \bibnamefont {Cordiner}}, \bibinfo {author} {\bibfnamefont {E.}~\bibnamefont {Herbst}}, \bibinfo {author} {\bibfnamefont {S.}~\bibnamefont {Kalenskii}},  \emph {et~al.},\ }\href@noop {} {\bibfield  {journal} {\bibinfo  {journal} {Science}\ }\textbf {\bibinfo {volume} {371}},\ \bibinfo {pages} {1265} (\bibinfo {year} {2021})}\BibitemShut {NoStop}%
\bibitem [{\citenamefont {Burkhardt}\ \emph {et~al.}(2021)\citenamefont {Burkhardt}, \citenamefont {Lee}, \citenamefont {Changala}, \citenamefont {Shingledecker}, \citenamefont {Cooke}, \citenamefont {Loomis}, \citenamefont {Wei}, \citenamefont {Charnley}, \citenamefont {Herbst}, \citenamefont {McCarthy} \emph {et~al.}}]{burkhardt2021discovery}%
  \BibitemOpen
  \bibfield  {author} {\bibinfo {author} {\bibfnamefont {A.~M.}\ \bibnamefont {Burkhardt}}, \bibinfo {author} {\bibfnamefont {K.~L.~K.}\ \bibnamefont {Lee}}, \bibinfo {author} {\bibfnamefont {P.~B.}\ \bibnamefont {Changala}}, \bibinfo {author} {\bibfnamefont {C.~N.}\ \bibnamefont {Shingledecker}}, \bibinfo {author} {\bibfnamefont {I.~R.}\ \bibnamefont {Cooke}}, \bibinfo {author} {\bibfnamefont {R.~A.}\ \bibnamefont {Loomis}}, \bibinfo {author} {\bibfnamefont {H.}~\bibnamefont {Wei}}, \bibinfo {author} {\bibfnamefont {S.~B.}\ \bibnamefont {Charnley}}, \bibinfo {author} {\bibfnamefont {E.}~\bibnamefont {Herbst}}, \bibinfo {author} {\bibfnamefont {M.~C.}\ \bibnamefont {McCarthy}},  \emph {et~al.},\ }\href@noop {} {\bibfield  {journal} {\bibinfo  {journal} {The Astrophysical Journal Letters}\ }\textbf {\bibinfo {volume} {913}},\ \bibinfo {pages} {L18} (\bibinfo {year} {2021})}\BibitemShut {NoStop}%
\bibitem [{\citenamefont {Sita}\ \emph {et~al.}(2022)\citenamefont {Sita}, \citenamefont {Changala}, \citenamefont {Xue}, \citenamefont {Burkhardt}, \citenamefont {Shingledecker}, \citenamefont {Lee}, \citenamefont {Loomis}, \citenamefont {Momjian}, \citenamefont {Siebert}, \citenamefont {Gupta} \emph {et~al.}}]{sita2022discovery}%
  \BibitemOpen
  \bibfield  {author} {\bibinfo {author} {\bibfnamefont {M.~L.}\ \bibnamefont {Sita}}, \bibinfo {author} {\bibfnamefont {P.~B.}\ \bibnamefont {Changala}}, \bibinfo {author} {\bibfnamefont {C.}~\bibnamefont {Xue}}, \bibinfo {author} {\bibfnamefont {A.~M.}\ \bibnamefont {Burkhardt}}, \bibinfo {author} {\bibfnamefont {C.~N.}\ \bibnamefont {Shingledecker}}, \bibinfo {author} {\bibfnamefont {K.~L.~K.}\ \bibnamefont {Lee}}, \bibinfo {author} {\bibfnamefont {R.~A.}\ \bibnamefont {Loomis}}, \bibinfo {author} {\bibfnamefont {E.}~\bibnamefont {Momjian}}, \bibinfo {author} {\bibfnamefont {M.~A.}\ \bibnamefont {Siebert}}, \bibinfo {author} {\bibfnamefont {D.}~\bibnamefont {Gupta}},  \emph {et~al.},\ }\href@noop {} {\bibfield  {journal} {\bibinfo  {journal} {The Astrophysical Journal Letters}\ }\textbf {\bibinfo {volume} {938}},\ \bibinfo {pages} {L12} (\bibinfo {year} {2022})}\BibitemShut {NoStop}%
\bibitem [{\citenamefont {Cernicharo}\ \emph {et~al.}(2024)\citenamefont {Cernicharo}, \citenamefont {Cabezas}, \citenamefont {Fuentetaja}, \citenamefont {Ag{\'u}ndez}, \citenamefont {Tercero}, \citenamefont {Janeiro}, \citenamefont {Juanes}, \citenamefont {Kaiser}, \citenamefont {Endo}, \citenamefont {Steber} \emph {et~al.}}]{cernicharo2024discovery}%
  \BibitemOpen
  \bibfield  {author} {\bibinfo {author} {\bibfnamefont {J.}~\bibnamefont {Cernicharo}}, \bibinfo {author} {\bibfnamefont {C.}~\bibnamefont {Cabezas}}, \bibinfo {author} {\bibfnamefont {R.}~\bibnamefont {Fuentetaja}}, \bibinfo {author} {\bibfnamefont {M.}~\bibnamefont {Ag{\'u}ndez}}, \bibinfo {author} {\bibfnamefont {B.}~\bibnamefont {Tercero}}, \bibinfo {author} {\bibfnamefont {J.}~\bibnamefont {Janeiro}}, \bibinfo {author} {\bibfnamefont {M.}~\bibnamefont {Juanes}}, \bibinfo {author} {\bibfnamefont {R.}~\bibnamefont {Kaiser}}, \bibinfo {author} {\bibfnamefont {Y.}~\bibnamefont {Endo}}, \bibinfo {author} {\bibfnamefont {A.}~\bibnamefont {Steber}},  \emph {et~al.},\ }\href@noop {} {\bibfield  {journal} {\bibinfo  {journal} {Astronomy \& Astrophysics}\ }\textbf {\bibinfo {volume} {690}},\ \bibinfo {pages} {L13} (\bibinfo {year} {2024})}\BibitemShut {NoStop}%
\bibitem [{\citenamefont {{Pilleri, P.}}\ \emph {et~al.}(2015)\citenamefont {{Pilleri, P.}}, \citenamefont {{Joblin, C.}}, \citenamefont {{Boulanger, F.}},\ and\ \citenamefont {{Onaka, T.}}}]{refId0}%
  \BibitemOpen
  \bibfield  {author} {\bibinfo {author} {\bibnamefont {{Pilleri, P.}}}, \bibinfo {author} {\bibnamefont {{Joblin, C.}}}, \bibinfo {author} {\bibnamefont {{Boulanger, F.}}}, \ and\ \bibinfo {author} {\bibnamefont {{Onaka, T.}}},\ }\href {\doibase 10.1051/0004-6361/201425590} {\bibfield  {journal} {\bibinfo  {journal} {A$\&$A}\ }\textbf {\bibinfo {volume} {577}},\ \bibinfo {pages} {A16} (\bibinfo {year} {2015})}\BibitemShut {NoStop}%
\bibitem [{\citenamefont {Kaiser}\ \emph {et~al.}(2015)\citenamefont {Kaiser}, \citenamefont {Parker},\ and\ \citenamefont {Mebel}}]{annurev:/content/journals/10.1146/annurev-physchem-040214-121502}%
  \BibitemOpen
  \bibfield  {author} {\bibinfo {author} {\bibfnamefont {R.~I.}\ \bibnamefont {Kaiser}}, \bibinfo {author} {\bibfnamefont {D.~S.}\ \bibnamefont {Parker}}, \ and\ \bibinfo {author} {\bibfnamefont {A.~M.}\ \bibnamefont {Mebel}},\ }\href {\doibase https://doi.org/10.1146/annurev-physchem-040214-121502} {\bibfield  {journal} {\bibinfo  {journal} {Annual Review of Physical Chemistry}\ }\textbf {\bibinfo {volume} {66}},\ \bibinfo {pages} {43} (\bibinfo {year} {2015})}\BibitemShut {NoStop}%
\bibitem [{\citenamefont {Cernicharo}\ \emph {et~al.}(2001)\citenamefont {Cernicharo}, \citenamefont {Heras}, \citenamefont {Tielens}, \citenamefont {Pardo}, \citenamefont {Herpin}, \citenamefont {Guélin},\ and\ \citenamefont {Waters}}]{Cernicharo_2001}%
  \BibitemOpen
  \bibfield  {author} {\bibinfo {author} {\bibfnamefont {J.}~\bibnamefont {Cernicharo}}, \bibinfo {author} {\bibfnamefont {A.~M.}\ \bibnamefont {Heras}}, \bibinfo {author} {\bibfnamefont {A.~G. G.~M.}\ \bibnamefont {Tielens}}, \bibinfo {author} {\bibfnamefont {J.~R.}\ \bibnamefont {Pardo}}, \bibinfo {author} {\bibfnamefont {F.}~\bibnamefont {Herpin}}, \bibinfo {author} {\bibfnamefont {M.}~\bibnamefont {Guélin}}, \ and\ \bibinfo {author} {\bibfnamefont {L.~B. F.~M.}\ \bibnamefont {Waters}},\ }\href {\doibase 10.1086/318871} {\bibfield  {journal} {\bibinfo  {journal} {The Astrophysical Journal}\ }\textbf {\bibinfo {volume} {546}},\ \bibinfo {pages} {L123} (\bibinfo {year} {2001})}\BibitemShut {NoStop}%
\bibitem [{\citenamefont {Cooke}\ \emph {et~al.}(2020)\citenamefont {Cooke}, \citenamefont {Gupta}, \citenamefont {Messinger},\ and\ \citenamefont {Sims}}]{Cooke_2020}%
  \BibitemOpen
  \bibfield  {author} {\bibinfo {author} {\bibfnamefont {I.~R.}\ \bibnamefont {Cooke}}, \bibinfo {author} {\bibfnamefont {D.}~\bibnamefont {Gupta}}, \bibinfo {author} {\bibfnamefont {J.~P.}\ \bibnamefont {Messinger}}, \ and\ \bibinfo {author} {\bibfnamefont {I.~R.}\ \bibnamefont {Sims}},\ }\href {\doibase 10.3847/2041-8213/ab7a9c} {\bibfield  {journal} {\bibinfo  {journal} {The Astrophysical Journal Letters}\ }\textbf {\bibinfo {volume} {891}},\ \bibinfo {pages} {L41} (\bibinfo {year} {2020})}\BibitemShut {NoStop}%
\bibitem [{\citenamefont {McGuire}\ \emph {et~al.}(2018)\citenamefont {McGuire}, \citenamefont {Burkhardt}, \citenamefont {Kalenskii}, \citenamefont {Shingledecker}, \citenamefont {Remijan}, \citenamefont {Herbst},\ and\ \citenamefont {McCarthy}}]{mcguire2018detection}%
  \BibitemOpen
  \bibfield  {author} {\bibinfo {author} {\bibfnamefont {B.~A.}\ \bibnamefont {McGuire}}, \bibinfo {author} {\bibfnamefont {A.~M.}\ \bibnamefont {Burkhardt}}, \bibinfo {author} {\bibfnamefont {S.}~\bibnamefont {Kalenskii}}, \bibinfo {author} {\bibfnamefont {C.~N.}\ \bibnamefont {Shingledecker}}, \bibinfo {author} {\bibfnamefont {A.~J.}\ \bibnamefont {Remijan}}, \bibinfo {author} {\bibfnamefont {E.}~\bibnamefont {Herbst}}, \ and\ \bibinfo {author} {\bibfnamefont {M.~C.}\ \bibnamefont {McCarthy}},\ }\href@noop {} {\bibfield  {journal} {\bibinfo  {journal} {Science}\ }\textbf {\bibinfo {volume} {359}},\ \bibinfo {pages} {202} (\bibinfo {year} {2018})}\BibitemShut {NoStop}%
\bibitem [{\citenamefont {McCarthy}\ \emph {et~al.}(2021)\citenamefont {McCarthy}, \citenamefont {Lee}, \citenamefont {Loomis}, \citenamefont {Burkhardt}, \citenamefont {Shingledecker}, \citenamefont {Charnley}, \citenamefont {Cordiner}, \citenamefont {Herbst}, \citenamefont {Kalenskii}, \citenamefont {Willis} \emph {et~al.}}]{mccarthy2021interstellar}%
  \BibitemOpen
  \bibfield  {author} {\bibinfo {author} {\bibfnamefont {M.~C.}\ \bibnamefont {McCarthy}}, \bibinfo {author} {\bibfnamefont {K.~L.~K.}\ \bibnamefont {Lee}}, \bibinfo {author} {\bibfnamefont {R.~A.}\ \bibnamefont {Loomis}}, \bibinfo {author} {\bibfnamefont {A.~M.}\ \bibnamefont {Burkhardt}}, \bibinfo {author} {\bibfnamefont {C.~N.}\ \bibnamefont {Shingledecker}}, \bibinfo {author} {\bibfnamefont {S.~B.}\ \bibnamefont {Charnley}}, \bibinfo {author} {\bibfnamefont {M.~A.}\ \bibnamefont {Cordiner}}, \bibinfo {author} {\bibfnamefont {E.}~\bibnamefont {Herbst}}, \bibinfo {author} {\bibfnamefont {S.}~\bibnamefont {Kalenskii}}, \bibinfo {author} {\bibfnamefont {E.~R.}\ \bibnamefont {Willis}},  \emph {et~al.},\ }\href@noop {} {\bibfield  {journal} {\bibinfo  {journal} {Nature Astronomy}\ }\textbf {\bibinfo {volume} {5}},\ \bibinfo {pages} {176} (\bibinfo {year} {2021})}\BibitemShut {NoStop}%
\bibitem [{\citenamefont {Lee}\ \emph {et~al.}(2021)\citenamefont {Lee}, \citenamefont {Changala}, \citenamefont {Loomis}, \citenamefont {Burkhardt}, \citenamefont {Xue}, \citenamefont {Cordiner}, \citenamefont {Charnley}, \citenamefont {McCarthy},\ and\ \citenamefont {McGuire}}]{lee2021interstellar}%
  \BibitemOpen
  \bibfield  {author} {\bibinfo {author} {\bibfnamefont {K.~L.~K.}\ \bibnamefont {Lee}}, \bibinfo {author} {\bibfnamefont {P.~B.}\ \bibnamefont {Changala}}, \bibinfo {author} {\bibfnamefont {R.~A.}\ \bibnamefont {Loomis}}, \bibinfo {author} {\bibfnamefont {A.~M.}\ \bibnamefont {Burkhardt}}, \bibinfo {author} {\bibfnamefont {C.}~\bibnamefont {Xue}}, \bibinfo {author} {\bibfnamefont {M.~A.}\ \bibnamefont {Cordiner}}, \bibinfo {author} {\bibfnamefont {S.~B.}\ \bibnamefont {Charnley}}, \bibinfo {author} {\bibfnamefont {M.~C.}\ \bibnamefont {McCarthy}}, \ and\ \bibinfo {author} {\bibfnamefont {B.~A.}\ \bibnamefont {McGuire}},\ }\href@noop {} {\bibfield  {journal} {\bibinfo  {journal} {The Astrophysical Journal Letters}\ }\textbf {\bibinfo {volume} {910}},\ \bibinfo {pages} {L2} (\bibinfo {year} {2021})}\BibitemShut {NoStop}%
\bibitem [{\citenamefont {Cernicharo}\ \emph {et~al.}(2021{\natexlab{a}})\citenamefont {Cernicharo}, \citenamefont {Ag{\'u}ndez}, \citenamefont {Cabezas}, \citenamefont {Tercero}, \citenamefont {Marcelino}, \citenamefont {Pardo},\ and\ \citenamefont {De~Vicente}}]{cernicharo2021pure}%
  \BibitemOpen
  \bibfield  {author} {\bibinfo {author} {\bibfnamefont {J.}~\bibnamefont {Cernicharo}}, \bibinfo {author} {\bibfnamefont {M.}~\bibnamefont {Ag{\'u}ndez}}, \bibinfo {author} {\bibfnamefont {C.}~\bibnamefont {Cabezas}}, \bibinfo {author} {\bibfnamefont {B.}~\bibnamefont {Tercero}}, \bibinfo {author} {\bibfnamefont {N.}~\bibnamefont {Marcelino}}, \bibinfo {author} {\bibfnamefont {J.~R.}\ \bibnamefont {Pardo}}, \ and\ \bibinfo {author} {\bibfnamefont {P.}~\bibnamefont {De~Vicente}},\ }\href@noop {} {\bibfield  {journal} {\bibinfo  {journal} {Astronomy \& Astrophysics}\ }\textbf {\bibinfo {volume} {649}},\ \bibinfo {pages} {L15} (\bibinfo {year} {2021}{\natexlab{a}})}\BibitemShut {NoStop}%
\bibitem [{\citenamefont {Cernicharo}\ \emph {et~al.}(2021{\natexlab{b}})\citenamefont {Cernicharo}, \citenamefont {Ag{\'u}ndez}, \citenamefont {Kaiser}, \citenamefont {Cabezas}, \citenamefont {Tercero}, \citenamefont {Marcelino}, \citenamefont {Pardo},\ and\ \citenamefont {De~Vicente}}]{cernicharo2021discovery}%
  \BibitemOpen
  \bibfield  {author} {\bibinfo {author} {\bibfnamefont {J.}~\bibnamefont {Cernicharo}}, \bibinfo {author} {\bibfnamefont {M.}~\bibnamefont {Ag{\'u}ndez}}, \bibinfo {author} {\bibfnamefont {R.~I.}\ \bibnamefont {Kaiser}}, \bibinfo {author} {\bibfnamefont {C.}~\bibnamefont {Cabezas}}, \bibinfo {author} {\bibfnamefont {B.}~\bibnamefont {Tercero}}, \bibinfo {author} {\bibfnamefont {N.}~\bibnamefont {Marcelino}}, \bibinfo {author} {\bibfnamefont {J.}~\bibnamefont {Pardo}}, \ and\ \bibinfo {author} {\bibfnamefont {P.}~\bibnamefont {De~Vicente}},\ }\href@noop {} {\bibfield  {journal} {\bibinfo  {journal} {Astronomy \& Astrophysics}\ }\textbf {\bibinfo {volume} {655}},\ \bibinfo {pages} {L1} (\bibinfo {year} {2021}{\natexlab{b}})}\BibitemShut {NoStop}%
\bibitem [{\citenamefont {Ben~Khalifa}\ and\ \citenamefont {Loreau}(2024)}]{ben2024rotational}%
  \BibitemOpen
  \bibfield  {author} {\bibinfo {author} {\bibfnamefont {M.}~\bibnamefont {Ben~Khalifa}}\ and\ \bibinfo {author} {\bibfnamefont {J.}~\bibnamefont {Loreau}},\ }\href@noop {} {\bibfield  {journal} {\bibinfo  {journal} {Monthly Notices of the Royal Astronomical Society}\ }\textbf {\bibinfo {volume} {527}},\ \bibinfo {pages} {846} (\bibinfo {year} {2024})}\BibitemShut {NoStop}%
\bibitem [{\citenamefont {Dzenis}\ \emph {et~al.}(2022)\citenamefont {Dzenis}, \citenamefont {Faure}, \citenamefont {McGuire}, \citenamefont {Remijan}, \citenamefont {Dagdigian}, \citenamefont {Rist}, \citenamefont {Dawes}, \citenamefont {Quintas-S{\'a}nchez}, \citenamefont {Lique},\ and\ \citenamefont {Hochlaf}}]{dzenis2022collisional}%
  \BibitemOpen
  \bibfield  {author} {\bibinfo {author} {\bibfnamefont {K.}~\bibnamefont {Dzenis}}, \bibinfo {author} {\bibfnamefont {A.}~\bibnamefont {Faure}}, \bibinfo {author} {\bibfnamefont {B.}~\bibnamefont {McGuire}}, \bibinfo {author} {\bibfnamefont {A.}~\bibnamefont {Remijan}}, \bibinfo {author} {\bibfnamefont {P.}~\bibnamefont {Dagdigian}}, \bibinfo {author} {\bibfnamefont {C.}~\bibnamefont {Rist}}, \bibinfo {author} {\bibfnamefont {R.}~\bibnamefont {Dawes}}, \bibinfo {author} {\bibfnamefont {E.}~\bibnamefont {Quintas-S{\'a}nchez}}, \bibinfo {author} {\bibfnamefont {F.}~\bibnamefont {Lique}}, \ and\ \bibinfo {author} {\bibfnamefont {M.}~\bibnamefont {Hochlaf}},\ }\href@noop {} {\bibfield  {journal} {\bibinfo  {journal} {The Astrophysical Journal}\ }\textbf {\bibinfo {volume} {926}},\ \bibinfo {pages} {3} (\bibinfo {year} {2022})}\BibitemShut {NoStop}%
\bibitem [{\citenamefont {Demes}\ \emph {et~al.}(2024)\citenamefont {Demes}, \citenamefont {Bop}, \citenamefont {Khalifa},\ and\ \citenamefont {Lique}}]{demes2024first}%
  \BibitemOpen
  \bibfield  {author} {\bibinfo {author} {\bibfnamefont {S.}~\bibnamefont {Demes}}, \bibinfo {author} {\bibfnamefont {C.~T.}\ \bibnamefont {Bop}}, \bibinfo {author} {\bibfnamefont {M.~B.}\ \bibnamefont {Khalifa}}, \ and\ \bibinfo {author} {\bibfnamefont {F.}~\bibnamefont {Lique}},\ }\href@noop {} {\bibfield  {journal} {\bibinfo  {journal} {Physical Chemistry Chemical Physics}\ }\textbf {\bibinfo {volume} {26}},\ \bibinfo {pages} {16829} (\bibinfo {year} {2024})}\BibitemShut {NoStop}%
\bibitem [{\citenamefont {Mandal}\ \emph {et~al.}(2022)\citenamefont {Mandal}, \citenamefont {Joy}, \citenamefont {Semenov},\ and\ \citenamefont {Babikov}}]{mandal2022mixed}%
  \BibitemOpen
  \bibfield  {author} {\bibinfo {author} {\bibfnamefont {B.}~\bibnamefont {Mandal}}, \bibinfo {author} {\bibfnamefont {C.}~\bibnamefont {Joy}}, \bibinfo {author} {\bibfnamefont {A.}~\bibnamefont {Semenov}}, \ and\ \bibinfo {author} {\bibfnamefont {D.}~\bibnamefont {Babikov}},\ }\href@noop {} {\bibfield  {journal} {\bibinfo  {journal} {ACS Earth and Space Chemistry}\ }\textbf {\bibinfo {volume} {6}},\ \bibinfo {pages} {521} (\bibinfo {year} {2022})}\BibitemShut {NoStop}%
\bibitem [{\citenamefont {Sakaizumi}\ \emph {et~al.}(1987)\citenamefont {Sakaizumi}, \citenamefont {Kikuchi}, \citenamefont {Ohashi},\ and\ \citenamefont {Yamaguchi}}]{sakaizumi1987microwave}%
  \BibitemOpen
  \bibfield  {author} {\bibinfo {author} {\bibfnamefont {T.}~\bibnamefont {Sakaizumi}}, \bibinfo {author} {\bibfnamefont {H.}~\bibnamefont {Kikuchi}}, \bibinfo {author} {\bibfnamefont {O.}~\bibnamefont {Ohashi}}, \ and\ \bibinfo {author} {\bibfnamefont {I.}~\bibnamefont {Yamaguchi}},\ }\href@noop {} {\bibfield  {journal} {\bibinfo  {journal} {Bulletin of the Chemical Society of Japan}\ }\textbf {\bibinfo {volume} {60}},\ \bibinfo {pages} {3903} (\bibinfo {year} {1987})}\BibitemShut {NoStop}%
\bibitem [{\citenamefont {Ford}\ and\ \citenamefont {Seitzman}(1978)}]{ford1978microwave}%
  \BibitemOpen
  \bibfield  {author} {\bibinfo {author} {\bibfnamefont {R.~G.}\ \bibnamefont {Ford}}\ and\ \bibinfo {author} {\bibfnamefont {H.~A.}\ \bibnamefont {Seitzman}},\ }\href@noop {} {\bibfield  {journal} {\bibinfo  {journal} {Journal of Molecular Spectroscopy}\ }\textbf {\bibinfo {volume} {69}},\ \bibinfo {pages} {326} (\bibinfo {year} {1978})}\BibitemShut {NoStop}%
\bibitem [{\citenamefont {Werner}\ \emph {et~al.}(2020)\citenamefont {Werner}, \citenamefont {Knowles}, \citenamefont {Manby}, \citenamefont {Black}, \citenamefont {Doll}, \citenamefont {Heßelmann}, \citenamefont {Kats}, \citenamefont {Köhn}, \citenamefont {Korona}, \citenamefont {Kreplin}, \citenamefont {Ma}, \citenamefont {Miller}, \citenamefont {Mitrushchenkov}, \citenamefont {Peterson}, \citenamefont {Polyak}, \citenamefont {Rauhut},\ and\ \citenamefont {Sibaev}}]{10.1063/5.0005081}%
  \BibitemOpen
  \bibfield  {author} {\bibinfo {author} {\bibfnamefont {H.-J.}\ \bibnamefont {Werner}}, \bibinfo {author} {\bibfnamefont {P.~J.}\ \bibnamefont {Knowles}}, \bibinfo {author} {\bibfnamefont {F.~R.}\ \bibnamefont {Manby}}, \bibinfo {author} {\bibfnamefont {J.~A.}\ \bibnamefont {Black}}, \bibinfo {author} {\bibfnamefont {K.}~\bibnamefont {Doll}}, \bibinfo {author} {\bibfnamefont {A.}~\bibnamefont {Heßelmann}}, \bibinfo {author} {\bibfnamefont {D.}~\bibnamefont {Kats}}, \bibinfo {author} {\bibfnamefont {A.}~\bibnamefont {Köhn}}, \bibinfo {author} {\bibfnamefont {T.}~\bibnamefont {Korona}}, \bibinfo {author} {\bibfnamefont {D.~A.}\ \bibnamefont {Kreplin}}, \bibinfo {author} {\bibfnamefont {Q.}~\bibnamefont {Ma}}, \bibinfo {author} {\bibfnamefont {I.}~\bibnamefont {Miller}, \bibfnamefont {Thomas~F.}}, \bibinfo {author} {\bibfnamefont {A.}~\bibnamefont {Mitrushchenkov}}, \bibinfo {author} {\bibfnamefont {K.~A.}\ \bibnamefont {Peterson}}, \bibinfo {author} {\bibfnamefont {I.}~\bibnamefont {Polyak}}, \bibinfo
  {author} {\bibfnamefont {G.}~\bibnamefont {Rauhut}}, \ and\ \bibinfo {author} {\bibfnamefont {M.}~\bibnamefont {Sibaev}},\ }\href {\doibase 10.1063/5.0005081} {\bibfield  {journal} {\bibinfo  {journal} {The Journal of Chemical Physics}\ }\textbf {\bibinfo {volume} {152}},\ \bibinfo {pages} {144107} (\bibinfo {year} {2020})},\ \Eprint {http://arxiv.org/abs/https://pubs.aip.org/aip/jcp/article-pdf/doi/10.1063/5.0005081/16680626/144107\_1\_online.pdf} {https://pubs.aip.org/aip/jcp/article-pdf/doi/10.1063/5.0005081/16680626/144107\_1\_online.pdf} \BibitemShut {NoStop}%
\bibitem [{\citenamefont {Boys}\ and\ \citenamefont {Bernardi}(1970)}]{boys1970calculation}%
  \BibitemOpen
  \bibfield  {author} {\bibinfo {author} {\bibfnamefont {S.~F.}\ \bibnamefont {Boys}}\ and\ \bibinfo {author} {\bibfnamefont {F.}~\bibnamefont {Bernardi}},\ }\href@noop {} {\bibfield  {journal} {\bibinfo  {journal} {Molecular physics}\ }\textbf {\bibinfo {volume} {19}},\ \bibinfo {pages} {553} (\bibinfo {year} {1970})}\BibitemShut {NoStop}%
\bibitem [{\citenamefont {Deegan}\ and\ \citenamefont {Knowles}(1994)}]{DEEGAN1994321}%
  \BibitemOpen
  \bibfield  {author} {\bibinfo {author} {\bibfnamefont {M.~J.}\ \bibnamefont {Deegan}}\ and\ \bibinfo {author} {\bibfnamefont {P.~J.}\ \bibnamefont {Knowles}},\ }\href {\doibase https://doi.org/10.1016/0009-2614(94)00815-9} {\bibfield  {journal} {\bibinfo  {journal} {Chemical Physics Letters}\ }\textbf {\bibinfo {volume} {227}},\ \bibinfo {pages} {321} (\bibinfo {year} {1994})}\BibitemShut {NoStop}%
\bibitem [{\citenamefont {Adler}\ \emph {et~al.}(2007)\citenamefont {Adler}, \citenamefont {Knizia},\ and\ \citenamefont {Werner}}]{adler2007simple}%
  \BibitemOpen
  \bibfield  {author} {\bibinfo {author} {\bibfnamefont {T.~B.}\ \bibnamefont {Adler}}, \bibinfo {author} {\bibfnamefont {G.}~\bibnamefont {Knizia}}, \ and\ \bibinfo {author} {\bibfnamefont {H.-J.}\ \bibnamefont {Werner}},\ }\href@noop {} {\bibfield  {journal} {\bibinfo  {journal} {The Journal of chemical physics}\ }\textbf {\bibinfo {volume} {127}} (\bibinfo {year} {2007})}\BibitemShut {NoStop}%
\bibitem [{\citenamefont {Dunning}(1989)}]{10.1063/1.456153}%
  \BibitemOpen
  \bibfield  {author} {\bibinfo {author} {\bibfnamefont {J.}~\bibnamefont {Dunning}, \bibfnamefont {Thom~H.}},\ }\href {\doibase 10.1063/1.456153} {\bibfield  {journal} {\bibinfo  {journal} {The Journal of Chemical Physics}\ }\textbf {\bibinfo {volume} {90}},\ \bibinfo {pages} {1007} (\bibinfo {year} {1989})},\ \Eprint {http://arxiv.org/abs/https://pubs.aip.org/aip/jcp/article-pdf/90/2/1007/18974738/1007\_1\_online.pdf} {https://pubs.aip.org/aip/jcp/article-pdf/90/2/1007/18974738/1007\_1\_online.pdf} \BibitemShut {NoStop}%
\bibitem [{\citenamefont {Knizia}\ \emph {et~al.}(2009)\citenamefont {Knizia}, \citenamefont {Adler},\ and\ \citenamefont {Werner}}]{knizia2009simplified}%
  \BibitemOpen
  \bibfield  {author} {\bibinfo {author} {\bibfnamefont {G.}~\bibnamefont {Knizia}}, \bibinfo {author} {\bibfnamefont {T.~B.}\ \bibnamefont {Adler}}, \ and\ \bibinfo {author} {\bibfnamefont {H.-J.}\ \bibnamefont {Werner}},\ }\href@noop {} {\bibfield  {journal} {\bibinfo  {journal} {The Journal of chemical physics}\ }\textbf {\bibinfo {volume} {130}} (\bibinfo {year} {2009})}\BibitemShut {NoStop}%
\bibitem [{\citenamefont {{Hutson}}\ and\ \citenamefont {{Green}}(1995)}]{molscat95}%
  \BibitemOpen
  \bibfield  {author} {\bibinfo {author} {\bibfnamefont {J.~M.}\ \bibnamefont {{Hutson}}}\ and\ \bibinfo {author} {\bibfnamefont {S.}~\bibnamefont {{Green}}},\ }\href@noop {} {\enquote {\bibinfo {title} {Molscat computer code, version 14, distributed by collaborative computational project 6},}\ }\bibinfo {howpublished} {Warington, UK: Daresbury Laboratory} (\bibinfo {year} {1995})\BibitemShut {NoStop}%
\bibitem [{\citenamefont {Schmitt}\ and\ \citenamefont {Meerts}(2018)}]{schmitt2018structures}%
  \BibitemOpen
  \bibfield  {author} {\bibinfo {author} {\bibfnamefont {M.}~\bibnamefont {Schmitt}}\ and\ \bibinfo {author} {\bibfnamefont {L.}~\bibnamefont {Meerts}},\ }in\ \href@noop {} {\emph {\bibinfo {booktitle} {Frontiers and advances in molecular spectroscopy}}}\ (\bibinfo  {publisher} {Elsevier},\ \bibinfo {year} {2018})\ pp.\ \bibinfo {pages} {143--193}\BibitemShut {NoStop}%
\bibitem [{\citenamefont {Flower}(2007)}]{flower2007molecular}%
  \BibitemOpen
  \bibfield  {author} {\bibinfo {author} {\bibfnamefont {D.}~\bibnamefont {Flower}},\ }\href@noop {} {\emph {\bibinfo {title} {Molecular collisions in the interstellar medium}}},\ Vol.~\bibinfo {volume} {42}\ (\bibinfo  {publisher} {Cambridge University Press},\ \bibinfo {year} {2007})\BibitemShut {NoStop}%
\bibitem [{\citenamefont {Arthurs}\ and\ \citenamefont {Dalgarno}(1960)}]{arthurs1960theory}%
  \BibitemOpen
  \bibfield  {author} {\bibinfo {author} {\bibfnamefont {A.}~\bibnamefont {Arthurs}}\ and\ \bibinfo {author} {\bibfnamefont {A.}~\bibnamefont {Dalgarno}},\ }\href@noop {} {\bibfield  {journal} {\bibinfo  {journal} {Proceedings of the Royal Society of London. Series A. Mathematical and Physical Sciences}\ }\textbf {\bibinfo {volume} {256}},\ \bibinfo {pages} {540} (\bibinfo {year} {1960})}\BibitemShut {NoStop}%
\bibitem [{\citenamefont {Green}(1976)}]{green1976rotational}%
  \BibitemOpen
  \bibfield  {author} {\bibinfo {author} {\bibfnamefont {S.}~\bibnamefont {Green}},\ }\href@noop {} {\bibfield  {journal} {\bibinfo  {journal} {The Journal of Chemical Physics}\ }\textbf {\bibinfo {volume} {64}},\ \bibinfo {pages} {3463} (\bibinfo {year} {1976})}\BibitemShut {NoStop}%
\bibitem [{\citenamefont {Garrison}\ \emph {et~al.}(1976)\citenamefont {Garrison}, \citenamefont {Lester},\ and\ \citenamefont {Miller}}]{garrison1976coupled}%
  \BibitemOpen
  \bibfield  {author} {\bibinfo {author} {\bibfnamefont {B.~J.}\ \bibnamefont {Garrison}}, \bibinfo {author} {\bibfnamefont {W.~A.}\ \bibnamefont {Lester}}, \ and\ \bibinfo {author} {\bibfnamefont {W.~H.}\ \bibnamefont {Miller}},\ }\href@noop {} {\bibfield  {journal} {\bibinfo  {journal} {The Journal of Chemical Physics}\ }\textbf {\bibinfo {volume} {65}},\ \bibinfo {pages} {2193} (\bibinfo {year} {1976})}\BibitemShut {NoStop}%
\bibitem [{\citenamefont {Manolopoulos}(1986)}]{manolopoulos1986improved}%
  \BibitemOpen
  \bibfield  {author} {\bibinfo {author} {\bibfnamefont {D.}~\bibnamefont {Manolopoulos}},\ }\href@noop {} {\bibfield  {journal} {\bibinfo  {journal} {The Journal of chemical physics}\ }\textbf {\bibinfo {volume} {85}},\ \bibinfo {pages} {6425} (\bibinfo {year} {1986})}\BibitemShut {NoStop}%
\bibitem [{\citenamefont {Faure}\ \emph {et~al.}(2019)\citenamefont {Faure}, \citenamefont {Dagdigian}, \citenamefont {Rist}, \citenamefont {Dawes}, \citenamefont {Quintas-S{\'a}nchez}, \citenamefont {Lique},\ and\ \citenamefont {Hochlaf}}]{faure2019interaction}%
  \BibitemOpen
  \bibfield  {author} {\bibinfo {author} {\bibfnamefont {A.}~\bibnamefont {Faure}}, \bibinfo {author} {\bibfnamefont {P.~J.}\ \bibnamefont {Dagdigian}}, \bibinfo {author} {\bibfnamefont {C.}~\bibnamefont {Rist}}, \bibinfo {author} {\bibfnamefont {R.}~\bibnamefont {Dawes}}, \bibinfo {author} {\bibfnamefont {E.}~\bibnamefont {Quintas-S{\'a}nchez}}, \bibinfo {author} {\bibfnamefont {F.}~\bibnamefont {Lique}}, \ and\ \bibinfo {author} {\bibfnamefont {M.}~\bibnamefont {Hochlaf}},\ }\href@noop {} {\bibfield  {journal} {\bibinfo  {journal} {ACS Earth and Space Chemistry}\ }\textbf {\bibinfo {volume} {3}},\ \bibinfo {pages} {964} (\bibinfo {year} {2019})}\BibitemShut {NoStop}%
\bibitem [{\citenamefont {Sogomonyan}\ \emph {et~al.}(2025)\citenamefont {Sogomonyan}, \citenamefont {Ben~Khalifa},\ and\ \citenamefont {Loreau}}]{sogomonyan2025rotational}%
  \BibitemOpen
  \bibfield  {author} {\bibinfo {author} {\bibfnamefont {K.}~\bibnamefont {Sogomonyan}}, \bibinfo {author} {\bibfnamefont {M.}~\bibnamefont {Ben~Khalifa}}, \ and\ \bibinfo {author} {\bibfnamefont {J.}~\bibnamefont {Loreau}},\ }\href@noop {} {\bibfield  {journal} {\bibinfo  {journal} {ACS Earth and Space Chemistry}\ } (\bibinfo {year} {2025})}\BibitemShut {NoStop}%
\bibitem [{\citenamefont {Ben~Khalifa}\ \emph {et~al.}(2022)\citenamefont {Ben~Khalifa}, \citenamefont {Dagdigian},\ and\ \citenamefont {Loreau}}]{ben2022interaction}%
  \BibitemOpen
  \bibfield  {author} {\bibinfo {author} {\bibfnamefont {M.}~\bibnamefont {Ben~Khalifa}}, \bibinfo {author} {\bibfnamefont {P.~J.}\ \bibnamefont {Dagdigian}}, \ and\ \bibinfo {author} {\bibfnamefont {J.}~\bibnamefont {Loreau}},\ }\href@noop {} {\bibfield  {journal} {\bibinfo  {journal} {The Journal of Physical Chemistry A}\ }\textbf {\bibinfo {volume} {126}},\ \bibinfo {pages} {9658} (\bibinfo {year} {2022})}\BibitemShut {NoStop}%
\bibitem [{\citenamefont {Ben~Khalifa}\ \emph {et~al.}(2023)\citenamefont {Ben~Khalifa}, \citenamefont {Dagdigian},\ and\ \citenamefont {Loreau}}]{ben2023collisional}%
  \BibitemOpen
  \bibfield  {author} {\bibinfo {author} {\bibfnamefont {M.}~\bibnamefont {Ben~Khalifa}}, \bibinfo {author} {\bibfnamefont {P.}~\bibnamefont {Dagdigian}}, \ and\ \bibinfo {author} {\bibfnamefont {J.}~\bibnamefont {Loreau}},\ }\href@noop {} {\bibfield  {journal} {\bibinfo  {journal} {Monthly Notices of the Royal Astronomical Society}\ }\textbf {\bibinfo {volume} {523}},\ \bibinfo {pages} {2577} (\bibinfo {year} {2023})}\BibitemShut {NoStop}%
\bibitem [{\citenamefont {Dubernet}\ \emph {et~al.}(2024)\citenamefont {Dubernet}, \citenamefont {Boursier}, \citenamefont {Denis-Alpizar}, \citenamefont {Ba}, \citenamefont {Moreau}, \citenamefont {Zw{\"o}lf}, \citenamefont {Amor}, \citenamefont {Babikov}, \citenamefont {Balakrishnan}, \citenamefont {Balan{\c{c}}a} \emph {et~al.}}]{dubernet2024basecol2023}%
  \BibitemOpen
  \bibfield  {author} {\bibinfo {author} {\bibfnamefont {M.}~\bibnamefont {Dubernet}}, \bibinfo {author} {\bibfnamefont {C.}~\bibnamefont {Boursier}}, \bibinfo {author} {\bibfnamefont {O.}~\bibnamefont {Denis-Alpizar}}, \bibinfo {author} {\bibfnamefont {Y.}~\bibnamefont {Ba}}, \bibinfo {author} {\bibfnamefont {N.}~\bibnamefont {Moreau}}, \bibinfo {author} {\bibfnamefont {C.}~\bibnamefont {Zw{\"o}lf}}, \bibinfo {author} {\bibfnamefont {M.}~\bibnamefont {Amor}}, \bibinfo {author} {\bibfnamefont {D.}~\bibnamefont {Babikov}}, \bibinfo {author} {\bibfnamefont {N.}~\bibnamefont {Balakrishnan}}, \bibinfo {author} {\bibfnamefont {C.}~\bibnamefont {Balan{\c{c}}a}},  \emph {et~al.},\ }\href@noop {} {\bibfield  {journal} {\bibinfo  {journal} {Astronomy \& Astrophysics}\ }\textbf {\bibinfo {volume} {683}},\ \bibinfo {pages} {A40} (\bibinfo {year} {2024})}\BibitemShut {NoStop}%
\end{thebibliography}%
\end{document}